\documentclass[preprints,article,galaxies,moreauthors,pdftex]{Definitions/mdpi} 
\firstpage{1} 
\pubvolume{1}
\issuenum{1}
\articlenumber{0}
\pubyear{2022}
\copyrightyear{2022}
\externaleditor{Academic Editor: Luigina Feretti} % please add
\datereceived{31 August 2022} 
\dateaccepted{29 November 2022} 
\datepublished{{date}} 
\hreflink{https://doi.org/} % If needed use \linebreak

\Title{Demonstration of Ultrawideband Polarimetry Using VLBI Exploration of Radio Astrometry (VERA)} % Please check intended meaning is retained.

\TitleCitation{ Demonstration of Ultrawideband Polarimetry Using VLBI Exploration of Radio Astrometry (VERA)}

\newcommand{\gbps}{Giga bit s$^{-1}\;$}
\Author{Yoshiaki Hagiwara $^{1,2,}$*$^{,\dagger}$\orcidA{}, Kazuhiro Hada $^{3,4}$\orcidB{}, Mieko Takamura $^{3,5}$\orcidC{}, Tomoaki Oyama $^{3}$\orcidD{}, Aya Yamauchi $^{3}$\orcidE{} \linebreak and Syunsaku Suzuki $^{6}$}
\AuthorNames{Yoshiaki Hagiwara, Kazuhiro Hada, Mieko Takamura, Tomoaki Oyama, Aya Yamauchi and Syunsaku Suzuki}
\AuthorCitation{Hagiwara, Y.; Hada, K.; Takamura, M.; Oyama, T.; Yamauchi, A.; Suzuki, S.}
\address{%
$^{1}$ \quad Natural Science Laboratory, Toyo University, 5-28-20 Hakusan Bunkyo-ku Tokyo, 112-8606%MDPI: Only city name should be in front of postcode. Please check: YH: Checked
, Japan\\
$^{2}$ \quad ASTRON and Joint Institute VLBI for ERIC (JIVE), Oude Hoogeveensedijk 4, 7991 PD Dwingeloo, \mbox{The Netherlands}\\
$^{3}$ \quad Mizusawa VLBI Observatory, National Astronomical Observatory of Japan, 2-12 Hoshigaoka, Mizusawa, Oshu, Iwate 023-0861, Japan; kazuhiro.hada@nao.ac.jp (K.H.); mieko.takamura@grad.nao.ac.jp (M.T.); t.oyama@nao.ac.jp (T.O.); a.yamauchi@nao.ac.jp (A.Y.) \\
$^{4}$ \quad Department of Astronomical Science, The Graduate University for Advanced Studies (SOKENDAI), 2-21-1 Osawa, Mitaka, Tokyo 181-8588, Japan \\
$^{5}$ \quad Department of Astronomy, Graduate School of Science, The University of Tokyo, 7-3-1, Hongo, Bunkyo, \mbox{Tokyo 113-0033,} Japan \\
$^{6}$ \quad National Astronomical Observatory of Japan, 2-21-1 Osawa, Mitaka, Tokyo 181-8588, Japan; syunsakujp2015@yahoo.co.jp 
}
\corres{Correspondence: yhagiwara@toyo.jp}
\firstnote{Current address: %MDPI: Current address shouldn't be duplicate with aff address. Please check and revise, YH: Current address is in Netherlands that is the same as my second aff.
 ASTRON and JIVE, Oude Hoogeveensedijk 4, 7991 PD Dwingeloo, The Netherlands} 
\abstract{We report on recent technical developments in the front- and back-ends for the four 20 m radio telescopes of the Japanese
Very-Long-Baseline Interferometry (VLBI) project, VLBI Exploration of Radio Astrometry (VERA).  We present a brief overview of a dual-circular polarization receiving and  ultrawideband (16 \gbps) recording systems that were installed on each of the four telescopes operating at 22 and 43 GHz bands. The wider-band capability improves the sensitivity of VLBI observations for continuum emission, and the dual-polarization capability enables the study of magnetic fields in relativistic jets ejected from supermassive black holes in active galactic nuclei and in sites of star formation and around evolved stars. We present the linear polarization intensity maps of extragalactic sources at 22 and 43 GHz obtained from the most recent test observations to show the state of the art of the VERA polarimetric observations.  At the end of this article, given the realization of VLBI polarimetry with VERA, we describe the future prospects for scientific aims and further technical developments. 
}
\keyword{polarization observations; radio interferometry; VLBI; VERA; AGN; magnetic field} 
\begin{document}
%%%%%%%%%%%%%%%%%%%%%%%%%%%%%%%%%%%%%%%%%%
%\setcounter{section}{-1} %% Remove this when starting to work on the template.
%\section{How to Use this Template}
%The template details the sections that can be used in a manuscript. Note that the order and names of article sections may differ from the requirements of the journal (e.g., the positioning of the Materials and Methods section). Please check the instructions on the authors' page of the journal to verify the correct order and names. For any questions, please contact the editorial office of the journal or support@mdpi.com. For LaTeX-related questions please contact latex@mdpi.com.%\endnote{This is an endnote.} % To use endnotes, please un-comment \printendnotes below (before References). Only journal Laws uses \footnote.
% The order of the section titles is: Introduction, Materials and Methods, Results, Discussion, Conclusions for these journals: aerospace,algorithms,antibodies,antioxidants,atmosphere,axioms,biomedicines,carbon,crystals,designs,diagnostics,environments,fermentation,fluids,forests,fractalfract,informatics,information,inventions,jfmk,jrfm,lubricants,neonatalscreening,neuroglia,particles,pharmaceutics,polymers,processes,technologies,viruses,vision
%
\nolinenumbers
\section{Introduction}
%
%{\bf \hl{Very-} %MDPI: Is the bold necessary? YH: Bold types in the manuscript are all removed.
Very-Long-Baseline Interferometry (VLBI) using radio telescopes spread across continents has been an important tool in radio astronomy, providing unprecedented angular resolution.  This can be exploited by imaging relativistic plasma jets ejected from supermassive black holes found in quasars and by mapping structures of cosmic masers associated with young stellar objects or evolved stars in the galaxy as well as megamasers in active galaxies.  VLBI astrometry of masers can obtain relative positional accuracy down to a $\sim$10 microarcsecond level, which in turn enables direct measurement of the rotation of our galaxy and constraints to be placed on the expansion rate of the universe and the dark-matter distribution (e.g., \citep{hon00, hagi09} and references therein).
%, for instance, mapping relativistic plasma jets ejected from super massive black holes found in quasars and structures of cosmic masers in massive or evolved stars and active galaxies.  VLBI has enabled the determinations of precise positions and motions of Galactic masers and quasars at 10 micro arcsecond level accuracy in order to measure the rotation of our Galaxy, the expansion rate of the Universe, and the dark matter distribution e.g.,\citep{hon00, hagi09} and references are therein. 
VLBI has long served as a geodetic technique to determine precise coordinates on the Earth, monitor Earth orientation parameters, and derive many other parameters of the Earth system (e.g., \citep{sch12,not17,gha21} and references therein). In order to improve accuracy and precision in the above measurements, it is necessary to record data at wider receiver bandwidths and higher recording rates.

\begin{figure}[H]
%\includegraphics[width=13.5 cm]{12GbpsSetup_OCTAD.pdf}
%\includegraphics[width=15 cm]{PolSetup_OCTAD.pdf}
%\begin{adjustwidth}{-\extralength}{0cm}

\includegraphics[angle=90, width=10.2cm,angle=270]{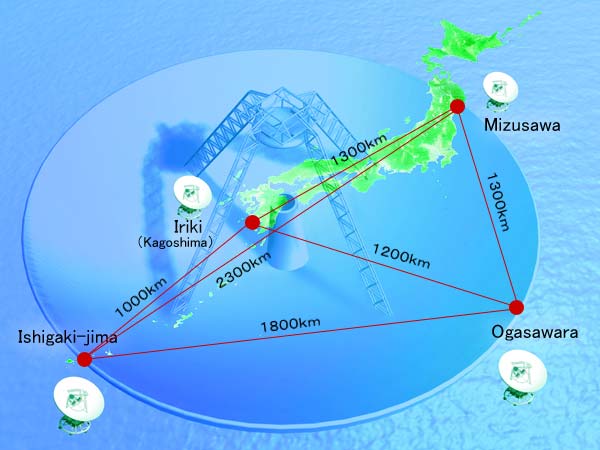}

\caption{An overview of the VERA array, consisting of four telescopes located at Mizusawa, Iriki, Ogasawara, and Ishigaki-Jima (Image credit:NAOJ) \label{fig1}}
\end{figure}

\textls[+15]{VLBI Exploration of Radio Astrometry (VERA) is a VLBI facility operated by the National Astronomical Observatory of Japan. It comprises four 20 m radio telescopes located across the country (Figure~\ref{fig1}). The basic specification of the telescopes and their performance characteristics are given in Table~\ref{tab1}. VERA was originally designed to reveal the three-dimensional structure of the Milky Way galaxy based on high-precision astrometry of galactic maser sources \cite{kawa00, hon00}. However, not only can VERA conduct high-precision galactic astrometry, but it can also be used for mapping compact extragalactic sources in radio wavelengths.}

The scientific capability of VERA has been recently improved by installing a new right-handed circular (RCP) polarization receiver for each of the 20 m telescopes that provides new dual-polarization capability across the array, new analog-to-digital samplers, and wideband disk-based recording systems that enable  data recording rates of 1, 2, 4, 8, and 16 \gbps \cite{oya12, oya16}. 

\textls[+15]{In this article, we describe the recent development and evaluation of the dual-polarization receiving capability at the front-ends of the 20 m telescopes. We present preliminary data and images obtained from the new system to illustrate the polarimetric observing performance.}
%comprehensible to scientists outside your particular field of research. Citing a journal paper \cite{ref-journal}. Now citing a book reference \cite{ref-book1,ref-book2} or other reference types \cite{ref-unpublish,ref-communication,ref-proceeding}. Please use the command \citep{ref-thesis,ref-url} for the following MDPI journals, which use author--date citation: Administrative Sciences, Arts, Econometrics, Economies, Genealogy, Humanities, IJFS, Journal of Intelligence, Journalism and Media, JRFM, Languages, Laws, Religions, Risks, Social Sciences, Literature.
%%%%%%%%%%%%%%%%%%%%%%%%%%%%%%%%%%%%%%%%%%
%

\begin{table}[H] 
\caption{Specifications of VERA 20 m telescopes  \label{tab1}}
\newcolumntype{C}{>{\centering\arraybackslash}X}

\begin{adjustwidth}{-\extralength}{0cm}
\centering 
\begin{tabularx}{\fulllength}{CC}
\toprule
\textbf{Items}	& \textbf{Values}\\
\midrule
Aperture efficiency \textsuperscript{1} 		& 45--50\% (22 GHz band), 35--50\% (43 GHz band)\\
Receiving frequency  \textsuperscript{1} & 20--24 GHz (22 GHz band),  42.2--44.5 GHz (43 GHz band) \\
Intermediate frequency &  4.7--7 GHz\\
%Receiver temperature (LNA)		& Data	\\
Polarizations  (22 and 43 GHz)&   Dual-circular polarization (LCP, RCP)   \\
%System temperature \textsuperscript{1}	& 	\\
%(K--band)	& 100--200 Kelvin		\\
%(Q-band)	& 180--250 Kelvin	\\
%Rec. rates		& 1,2,4,8,16 Giga bit sec$^{-1}$		\\
%Correlation	& FX-type, software correlator (softcos)		\\
\bottomrule
\end{tabularx}
\end{adjustwidth}
\noindent{\footnotesize{\textsuperscript{1} \citep{vera19}}}
\end{table}

\vspace{-5pt}

\vspace{-5pt} 

\section{Instrumentation}
\subsection{Overview}
\begin{figure}[H]

\includegraphics[width=17cm,angle=270]{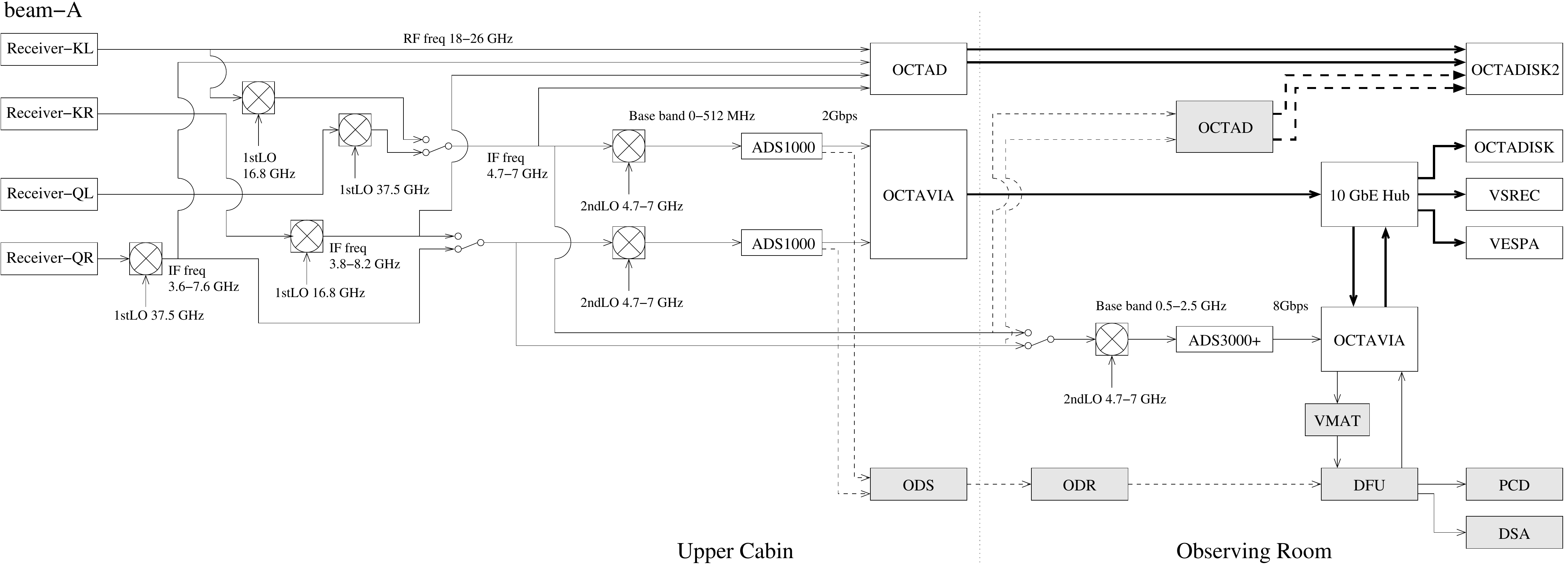}
%\end{adjustwidth}
%
\caption{Schematic diagram of the front-end part of the 22 GHz and 43 GHz receivers (beam-A) on the 20 m telescopes of VERA. 
%Receiver-KL indicates a 22 GHz band LCP receiver, and Receiver-KR does a 22 GHz RCP one.  Receiver-QL and -QR indicate a 43 GHz band LCP and a 43 GHz RCP receiver, respectively. 
The two-letter receiver abbreviations encode frequency band in the first letter (K for 22 GHz, Q for 43 GHz) and polarization in the second letter.
Signal paths follow the arrows. The 22 GHz radio frequency (RF) signals (18--26 GHz) are directly converted to intermediate frequency (IF) signals (4.7--7 GHz) using a "direct" sampler OCTAD. The 43 GHz RF signals are converted via the frequency down-converter to IF signals.  \label{fig2}}
\end{figure}

%
%We consider to separate the receiver system to the front-end for capturing and amplifying radio signals with low-noise amplifiers (LNA), the frequency down-converters, and intermediate frequency (IF), and the back-end where signals are converted to video frequency range and recorded to digital signal processing instruments. \\
Here, we separately consider the front- and back-end systems.  The former comprises capturing and amplifying radio signals with low-noise amplifiers (LNA), the frequency down-conversion, and intermediate frequency (IF) channels; the latter comprises the conversion of the IF signals to video-frequency range and the subsequent recording to digital signal processing instruments.  
The telescope front-end, IF, and the back-end system are shown in Figure~\ref{fig2}.
%%% RMC: does this sentence about the new RCP Rx belong in the "Front End" sub-section(?)  I've merged it with the existing sentence there...
%New receivers at right-handed circular polarization (RCP) are installed on the 20 m telescopes at each  of VERA, which enables to conduct dual polarization observations with the existing left-handed circular (LCP) receivers at all four s of VERA. 

\subsection{Front-End}
VERA is equipped with a unique dual-beam receiver system (termed beam-A and beam-B) at both 22 and 43 GHz bands. Each of the two beams can observe a different source, with a separation between the two sources up to 2 degrees in the sky \cite{kawa00}. 
Since the construction of the VERA, the 20m telescopes have been equipped with only a left-handed circular polarization (LCP) receiver for each of the two beams. 
%{\bf It is thus necessary} that a receiver with right-handed circular polarization (RCP) is installed in the frond-end on the all telescopes. 
New right-handed circular polarization (RCP) receivers were installed in the front-end of all the telescopes, which combined with the already present LCP receivers enables for the first time dual-polarization observations by VERA.
In this development, the RCP receivers are installed only for the beam-A at each telescope but not yet for the beam-B.

 In Figure~\ref{fig2}, the signal paths for the two feed horns at each of the two frequency bands are shown, illustrative of various stages of mixing/switching within the receiver stage. 
 Radio frequency (RF) signals received at the feed horn at each of the two frequency bands are divided into left- or right-handed circular polarizations by a circular polarizer, and then amplified by the LNA in the Cryogenic Dewar. Pictures of the construction inside the Dewar are shown in Figure~\ref{fig3}. The amplified signals in each circular polarization are converted to intermediate frequency (IF) via frequency down-converters appropriate to each of the frequency bands.

\vspace{-3pt}

\begin{figure}[H]

\includegraphics[width=13cm]{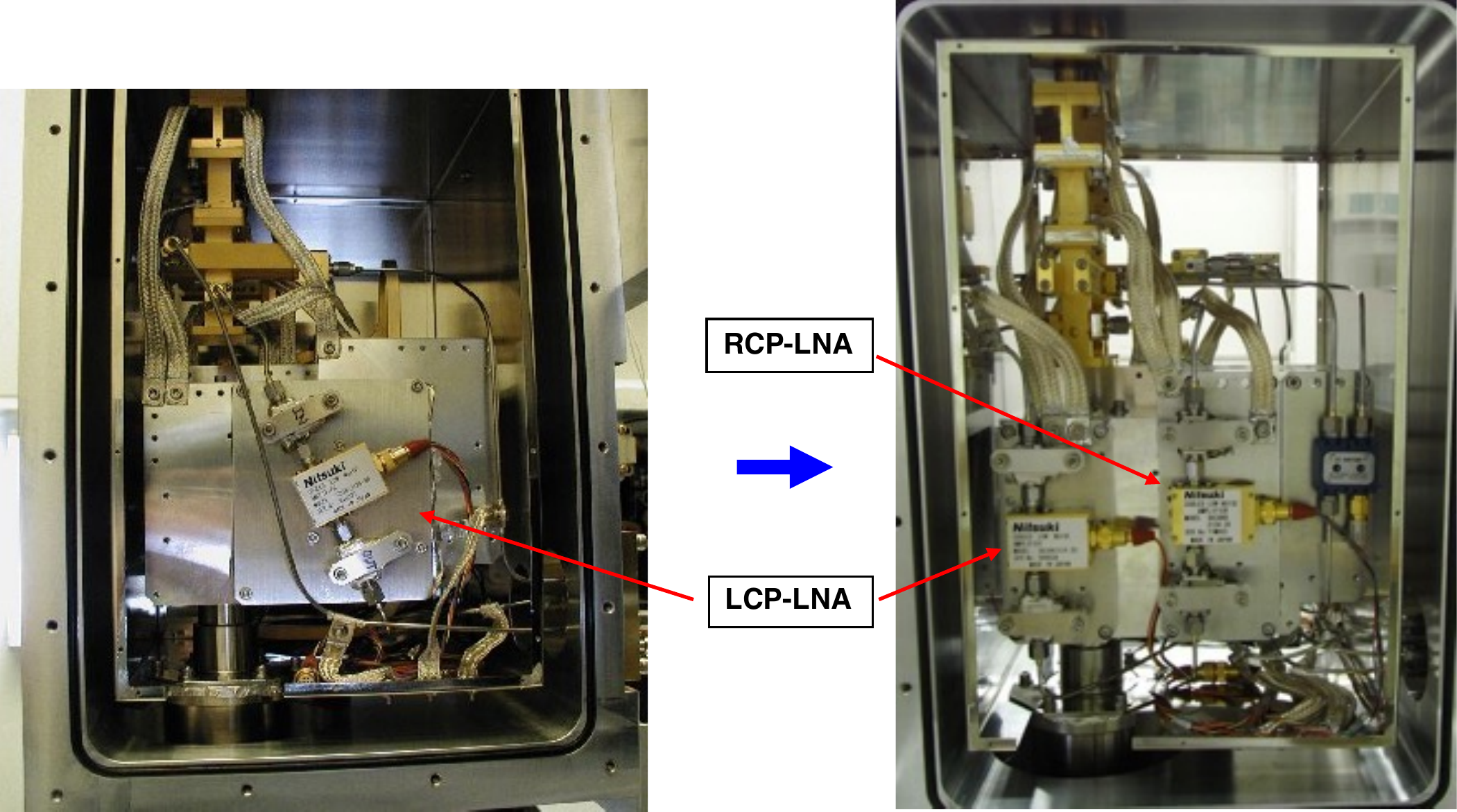} 
%\vspace{5mm}
%\vspace{9mm}
\includegraphics[width=13cm]{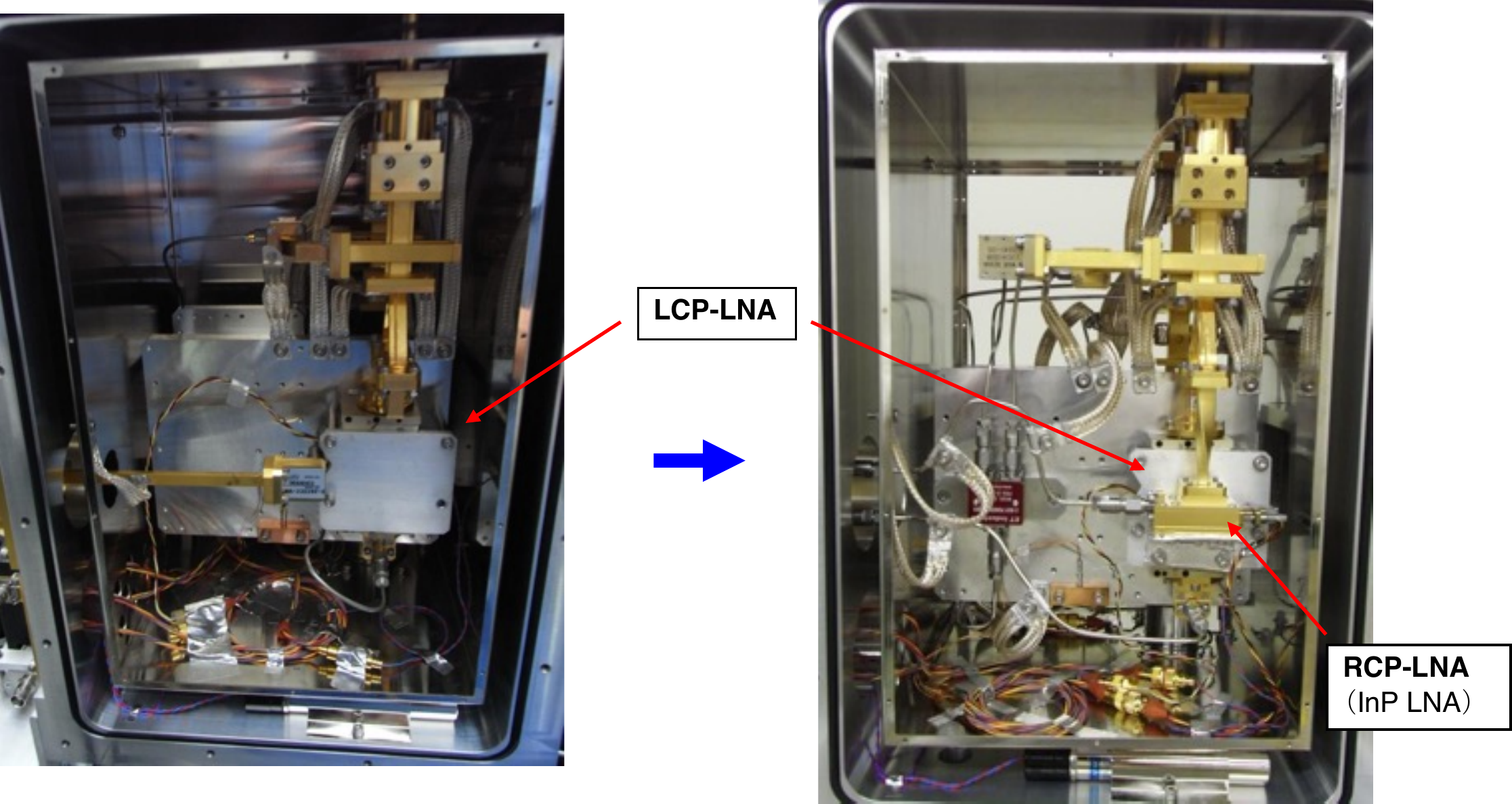}
%\vspace{3mm}

%\end{adjustwidth}
\caption{\textit{Cont}.\label{fig3}}
\end{figure}   
%%%%

%
%Figure 3
\begin{figure}[H]\ContinuedFloat

%\vspace{3mm}
\includegraphics[width=13cm]{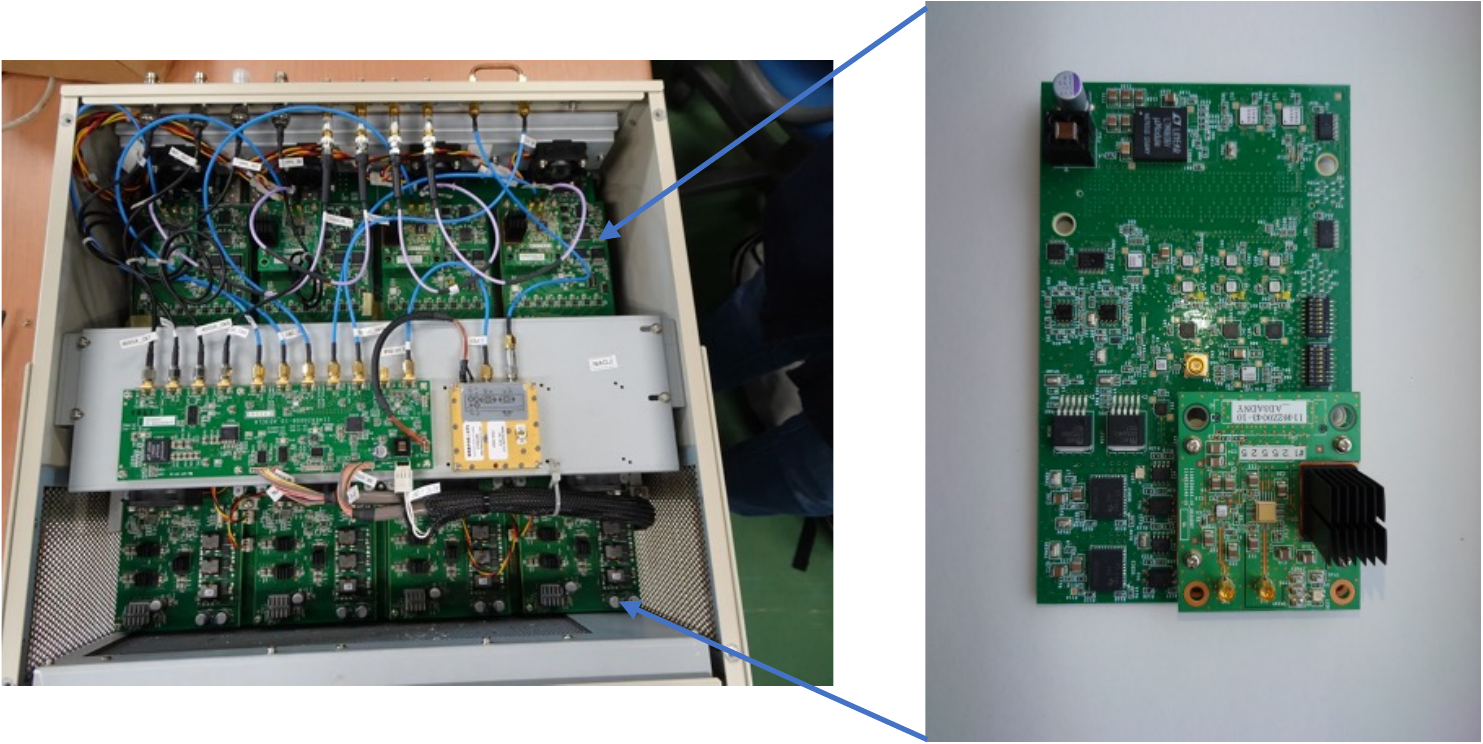}
%\end{adjustwidth}
\caption{Interiors of the cryogenic dewars with the single-polarization (LCP) 22 GHz (\textbf{top left)} and 43 GHz band (\textbf{middle left}) receivers, before 2018. Interiors of the dewars with the new dual-circular polarization at 22 GHz (\textbf{top right)} and 43 GHz (\textbf{middle right)} band receivers  (photos by Japan Communication Equipment Co., Ltd.).\textbf{(Bottom left)}: an interior of OCTAD and \textbf{(bottom right)}: an OCTAD channel board mounted on an FPGA board \label{fig3}}
\end{figure}   
%%%%

\subsection{Intermediate Frequency (IF) and Back-End}
\textls[+25]{A new analog-to-digital sampler (OCTAD: Optically Connected Array for VLBI Exploration Analog-to-Digital Converter) that enables frequency down-conversion at 18--26 GHz has been installed on each of the telescopes, which enables VLBI data recording at the rates of 4, 8, and 16 \gbps for each of the two polarizations, or up to 32 \gbps combined over both polarizations. At the time of mid-2022, the highest recording rate at the back-end is 16 \gbps. The details of the broad-band observing system for VERA will be found in a forthcoming paper (Oyama et al. \mbox{in preparation).}}

The 22 GHz band RF signals from the receivers are transmitted to through the intermediate frequency (4.7--7.0 GHz) using OCTAD, a "direct" sampler, while the 43 GHz band signals are converted to IF using the 43 GHz band down-converter, since OCTAD is not capable of directly sampling signals at frequencies higher than the 22 GHz band. Note, however, that in the observations described here, the frequency conversion at the 22 GHz band was conducted by the frequency converter rather than by OCTAD. 
With installations of both the new RCP receiver and the new back-end system, ultrawideband VLBI polarimetry is realized using VERA. % Please check intended meaning is retained.  YH: Confirmed.
\section{Observations}
%
%\subsection{Wide-band polarimetric Observations}
%
We performed commissioning wideband polarimetric observations with VERA on 2022 March 8 at 22 GHz band and on March 13 at 43 GHz band. Each session lasted 19 hours in a continuous observing track, in which we observed a number of radio sources, as listed in Table~\ref{tab2}. These observations were part of a VERA wideband science program on narrow-line Seyfert 1 galaxies, but we could also make use of bright calibrators in these observations to examine the polarization performance of VERA. While a dedicated study on the science targets will be published in a separate paper (Takamura et al. in preparation), here, we focus on the analysis of a few bright calibrators to demonstrate the polarimetric VLBI performance of VERA.  

All four VERA telescopes participated in good weather conditions throughout the sessions, with typical system temperatures of 100--200 K at the 22 GHz band and \mbox{200--800 K} at 43 GHz (Table~\ref{tab2}). Both sessions were performed at a recording rate of 16 \gbps, where signals were received in four 512\,MHz sub-bands in two circular polarizations. Thanks to the long observing tracks, most of the sources were observed over a wide range of parallactic angles, which was important to perform reliable calibration of %MDPI: Is the bold necessary?  YH: No, it's removed.
 the polarization leakage terms (D-terms). Correlation of the data was executed using the Mizusawa Software Correlator (softcos), an FX-type software correlator at NAOJ Mizusawa Observatory, which is also known as OCTACOR2. More details of the observations are summarized in Table~\ref{tab2}. 
 
 Figure~\ref{fig4} displays the time series of delay solutions from a global fringe fit for a portion of each session. % Please check intended meaning is retained. YH: Yes. Confirmed
%Figure~\ref{fig4} displays the time series of delay solutions from a global fringe fit for a portion of the sessions. 
%%%%%%bob
Uncertainties in each of the solutions are typically 2--3 nanoseconds;
systematic and calibration errors were considered in a standard way, and fitting errors that occur when obtaining delay solutions were also taken into account.

\vspace{-3pt}
\vspace{-3pt}   

\begin{table}[H] 
\caption{Summary of observations  \label{tab2}}

\begin{adjustwidth}{-\extralength}{0cm}
\centering 
\setlength{\cellWidtha}{\fulllength/3-2\tabcolsep-0.2in}
\setlength{\cellWidthb}{\fulllength/3-2\tabcolsep+0.1in}
\setlength{\cellWidthc}{\fulllength/3-2\tabcolsep+0.1in}
\scalebox{1}[1]{\begin{tabularx}{\fulllength}{
>{\PreserveBackslash\centering}m{\cellWidtha}
>{\PreserveBackslash\centering}m{\cellWidthb}
>{\PreserveBackslash\centering}m{\cellWidthc}}
\toprule
\textbf{Items}	& \textbf{22 GHz}& \textbf{43 GHz}\\
\midrule
Dates	& 8 March 2022 & 13 March 2022\\
Duration  (hours) & 19 &19\\
Observed sources \\
(targets)  &0321+340, 0846+513, 0946+006, 1219+044, 1505+03, J2118-07 &0321+340, 0846+513, 0946+006, 1219+044, 1505+03, J2118-07  \\ \\
(calibrators) & 0235+164, 0528+134, 3C 273,  3C 454.3, 3C 84,  OJ 287, W3OH & 0235+164, 0528+134, 3C 273,  3C 454.3, 3C 84,   OJ 287, Orion-KL \\
System temperatures \textsuperscript{1} (K) & & \\
Mizusawa, Iriki&  100-200(LCP), 100-200(RCP) &  200-800(LCP), 200-1000(RCP)\\
Ogasawara, Ishigaki&  150-300(LCP), 150-400(RCP) & 200-800(LCP), 200-1000(RCP) \\
The number of the sub-band	& 4 & 4 \\
IF sub-bandwidth \textsuperscript{2}	& 512 MHz & 512 MHz \\
%(K--band) &  0235+164, 0321+340, 0528+134, 0846+513, 0946+006, 1219+044, 3C 273,  3C 454.3, 3C 84, J 1505+03, J 2118-07,  OJ 287, W3OH\\
% (Q--band) &  0235+164, 0321+340, 0528+134, 0846+513, 0946+006, 1219+044, 3C 273,  3C 454.3, 3C 84, J 1505+03, J 2118-07,  OJ 287, Orion-KL
%The number of telescopes & 4&4 \\
%The number of circular polarizations & 2 &2\\	
Center frequencies of each \mbox{sub-band (GHz)} &21.715, 22.227, 22.739, 23.251  &42.682,  43.194, 43.706, 44.218\\
%(K--band) & 21.715, 22.227, 22.739, 23.251 GHz\\
% (Q--band) &42.682,  43.194, 43.706, 44.218 GHz\\
Recording rates		& 16 Gigabit sec$^{-1}$ &	16 Gigabit sec$^{-1}$	\\
Correlation	& FX-type, software correlator (softcos) &	softcos	\\
\bottomrule
\end{tabularx}}
\end{adjustwidth}
\noindent{\footnotesize{\textsuperscript{1} Typical values during observations depending on each of the four telescopes.}}\\
\noindent{\footnotesize{\textsuperscript{2} The total bandwidth is  2048 MHz (4 $\times$ 512 MHz) per  polarization.}}
\end{table}
\unskip
%
%Figure5

\begin{figure}[H]
%\centering
%\begin{adjustwidth}{-\extralength}{0cm}
\label{fig4}
\includegraphics[width=9.5cm]{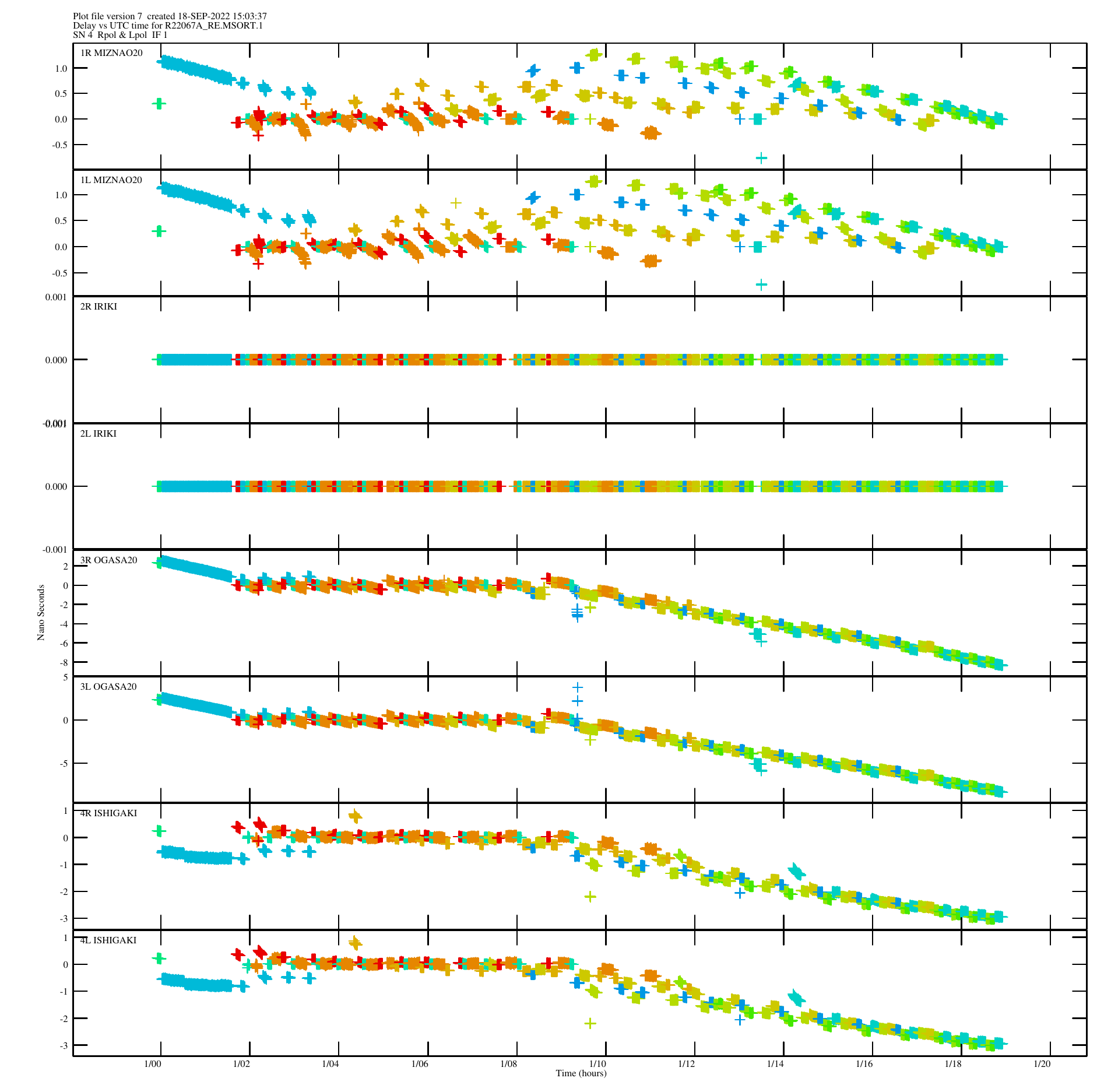}

%\end{adjustwidth}
\caption{\textit{Cont}.}
\end{figure}

\begin{figure}[H]\ContinuedFloat
%\centering
%\begin{adjustwidth}{-\extralength}{0cm}

\includegraphics[width=9.5cm]{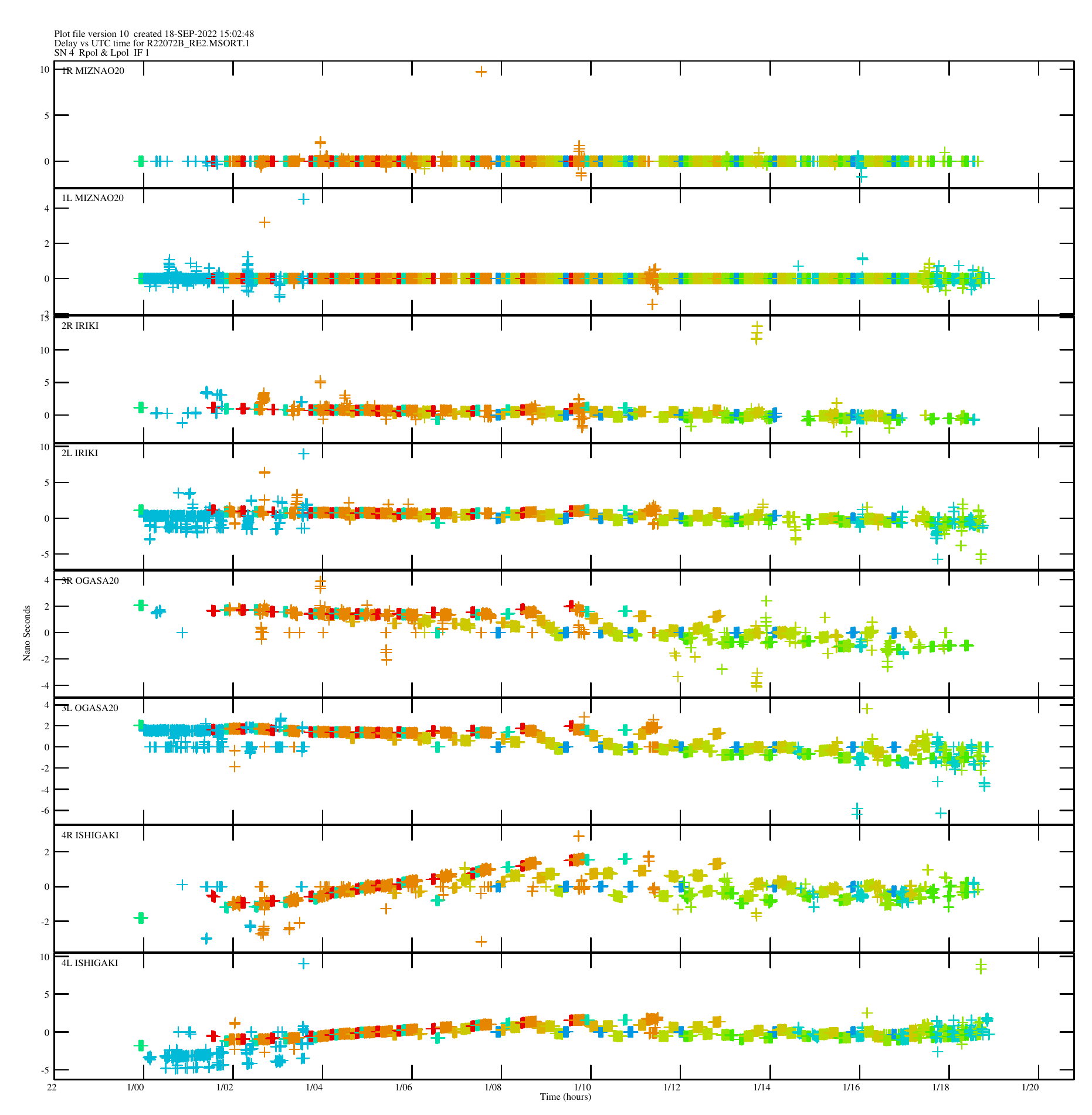}
%\end{adjustwidth}
\caption{Plots %MDPI: Please change the hyphen (-) into minus sign (−, “U+2212”). e.g., “-1” should be “−1”.
 of the delay solutions obtained at 22 GHz (\textbf{upper}) and 43 GHz (\textbf{lower}), with each point being indicated in different colors that correspond to different sources. Units of vertical and horizontal axes are in nanoseconds and hours, respectively.  \label{fig4}}
\end{figure}   
%Errors of each point are about 10\%.
%In our experiments, both K- and Q-band signals were received with dual-circular polarization.
%To conduct accurate polarization calibration, the instrumental polarization should be determined by observations of a bright calibrator source  over a wide range in parallactic angle and observations of an unpolarized source is necessary. Nine calibrator sources were observed at the both 22 and 43 GHz bands including a non-polarized source and sources with some structures. The observed sources and some observing parameters at each frequency band are listed in table \ref{tab2}. \\
%
%These experiments were observed at K- and Q-bands with 4 stations each recording data at a rate of 16 Gigabit s$^{-1}$: 2-bit Nyquist sampling of 4 $\times$ 512 MHz of bandwidth in two circular polarizations.
% where it moves through enough parallactic angle during the observation 
%the bandpass calibration was made using the bright quasar, 
%

%Correlation of the data was executed using the Mizusawa Software Correlator (softcos), an FX-type software correlator at NAOJ Mizusawa Observatory, which is also known as  OCTACOR2.
%After the correlation, the data were converted to FITS file format and data analysis was carried out using the Astronomical
%Image Processing System (AIPS) developed by the National Radio Astronomy Observatory (NRAO).
%
\section{Data Reduction}
The initial data calibration was performed with the Astronomical Image Processing System (AIPS) developed at the National Radio Astronomy Observatory (NRAO) \cite{gre90}. Since the polarimetric observations of VERA were performed with a single-beam mode, the data reduction procedures of VERA polarimetric data were similar to those of other VLBI networks, such as the Very-Long-Baseline Array (VLBA) and the Korean VLBI Network (KVN).  CLEAN imaging of Stokes I, Q, and U were performed in DIFMAP \cite{she97}. For the calibration of antenna gains, self-calibration of both phase and amplitude was performed using the Stokes I data set. D-terms were solved for each 512\,MHz sub-band separately by using the AIPS task LPCAL.

Note that corrections of absolute EVPA were not performed in our present analysis, since that requires near-in-time absolute EVPA information monitored by external (single-dish or connected array) facilities.  % Please check intended meaning is retained. YH : Confirmed.
\section{Results and Verification}
\subsection{Overview}
The new RCP receivers were installed and tested at all four telescopes in March 2022. In this section, we explain the procedures taken to test and verify the new system during the test observations.
\subsection{D-Terms}

\begin{figure}[H]
%\begin{adjustwidth}{-\extralength}{0cm}

\includegraphics[width=11cm]{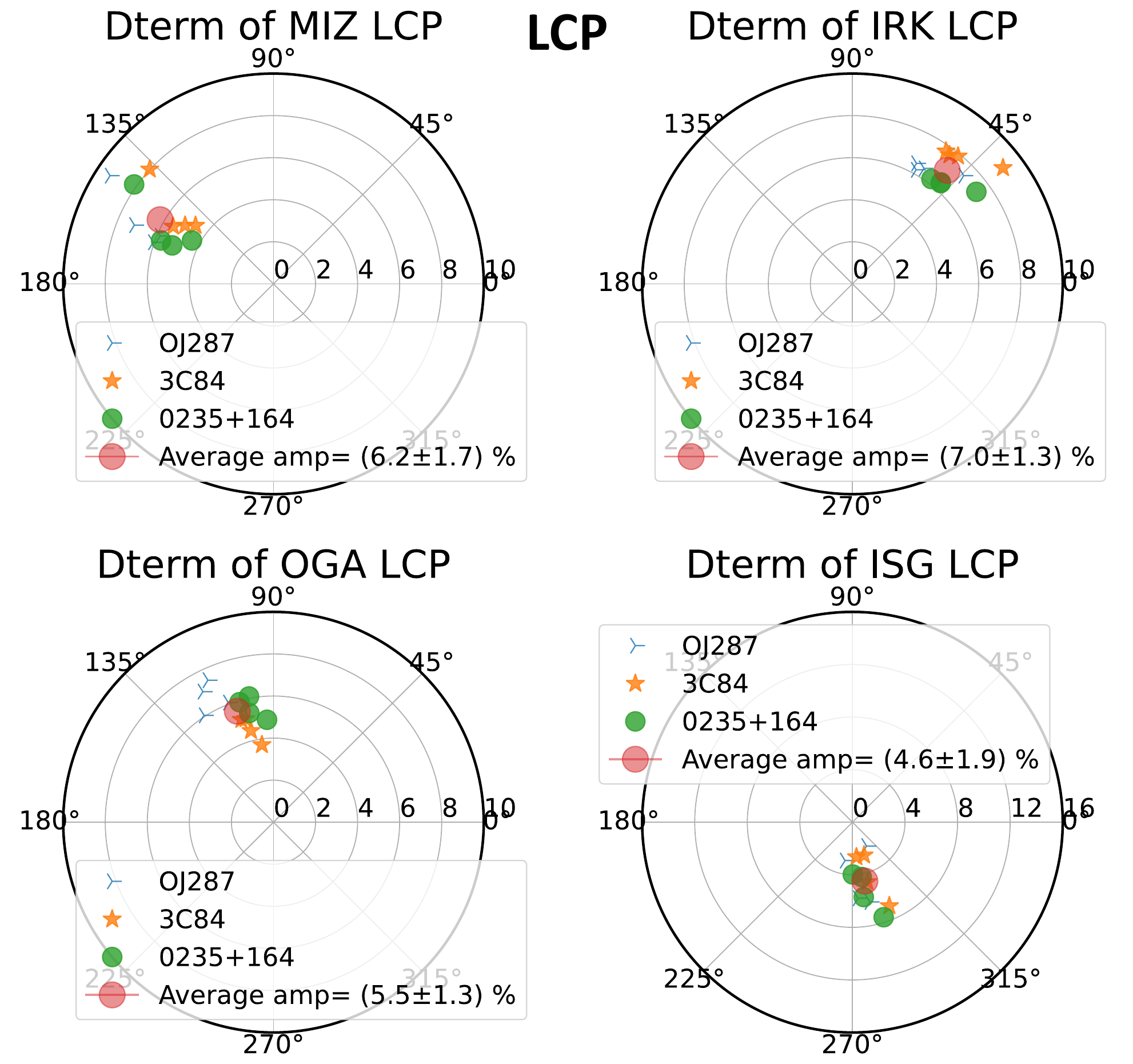}
\includegraphics[width=11cm]{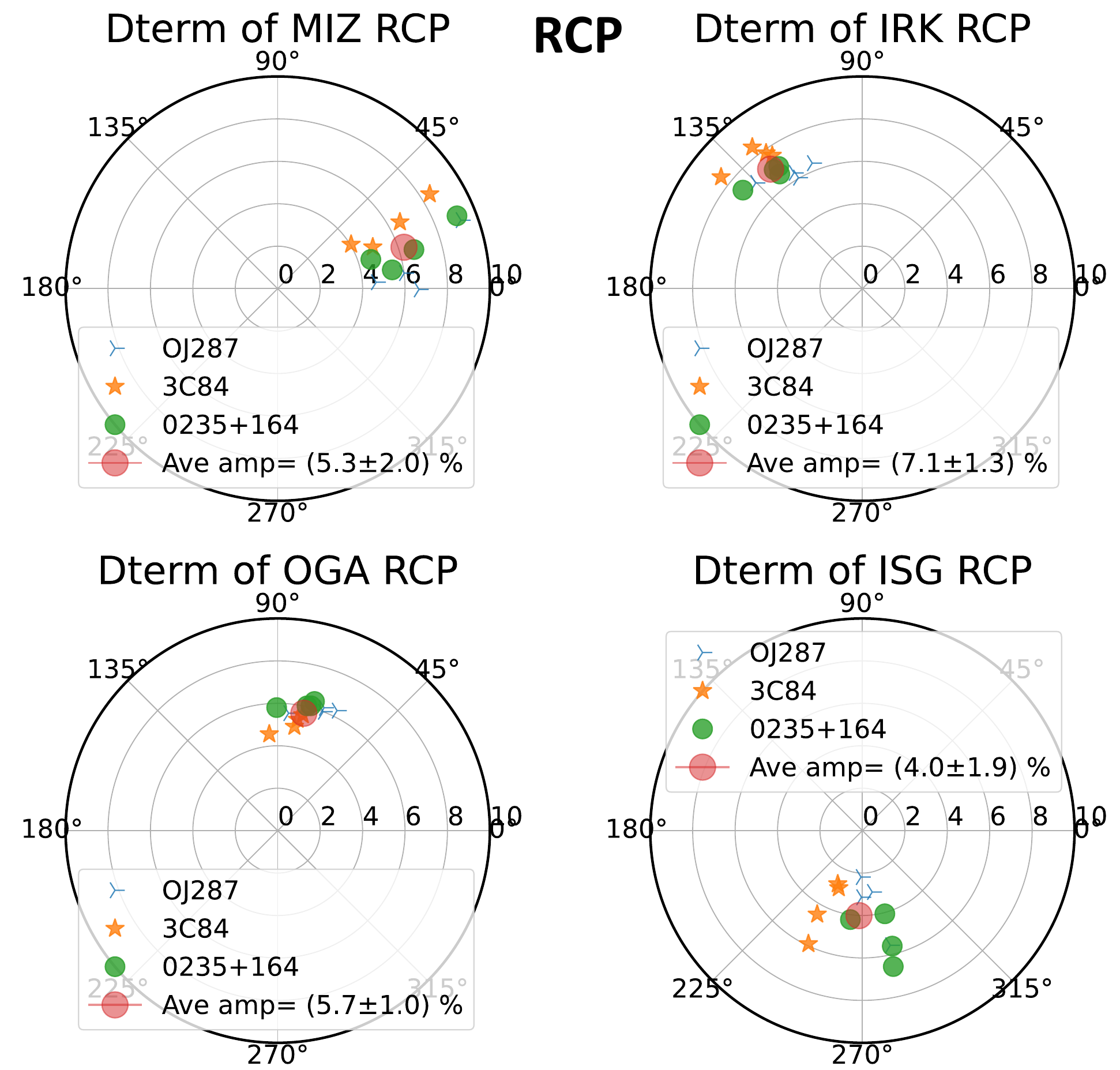}
%\end{adjustwidth}
\caption{{D-terms} %MDPI:  Please check the overlapped content
 obtained using different calibrator sources (different colors)
at 22 GHz observations by combining all four IF sub-bands. Each panel shows D-terms of each of the four VERA 20 m telescopes at each circular polarization (\textbf{upper}: LCP, \textbf{lower}: RCP). D-terms obtained by averaging over the sources are also shown. The radius of each plot is in percent.\label{fig5}} % Please check intended meaning is retained.
\end{figure}

\begin{figure}[H]
%\begin{adjustwidth}{-\extralength}{0cm}

%\includegraphics[width=10cm]{dterm-q-4IF_LCP.pdf}
\includegraphics[width=11cm]{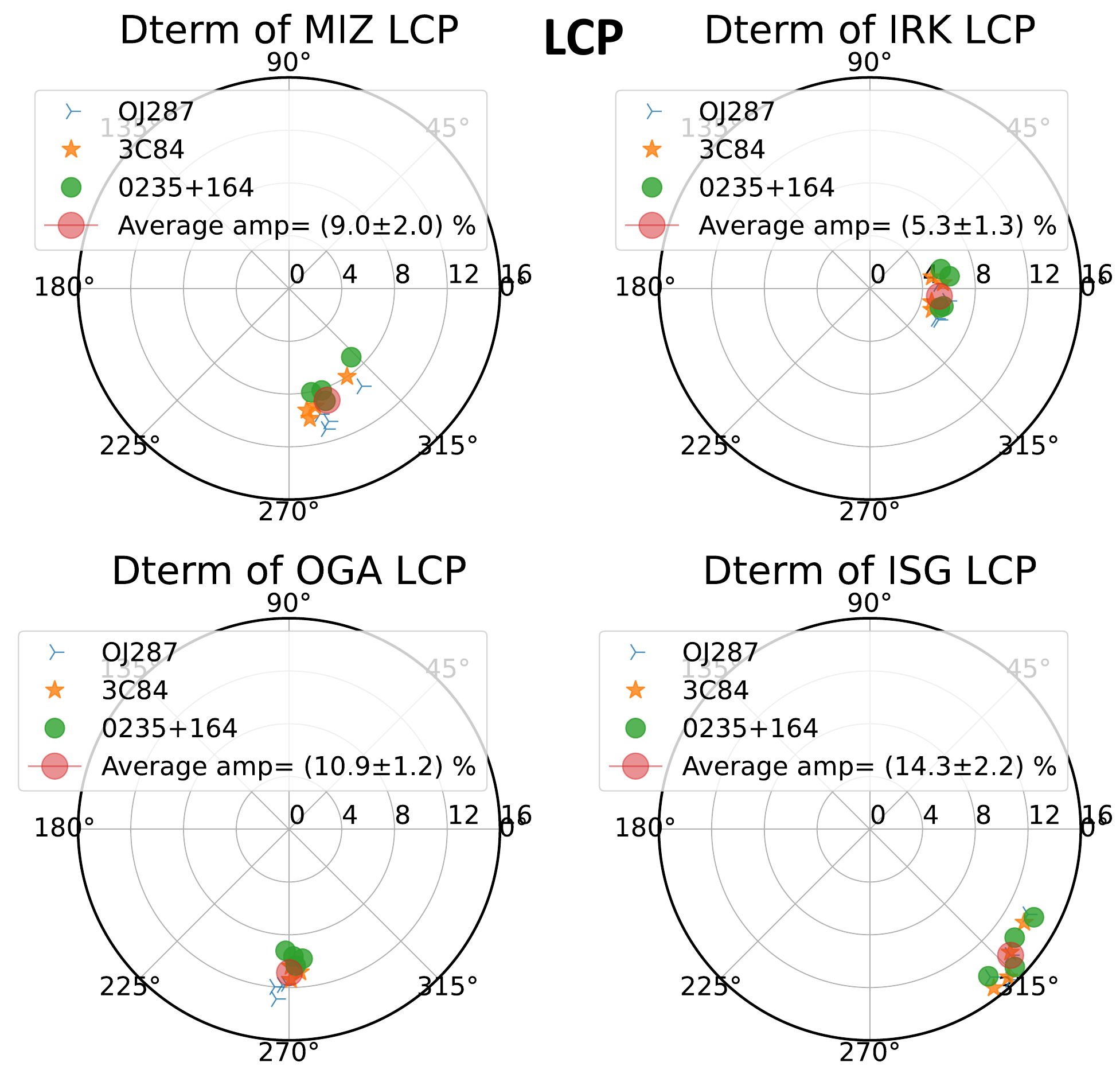}
\includegraphics[width=11cm]{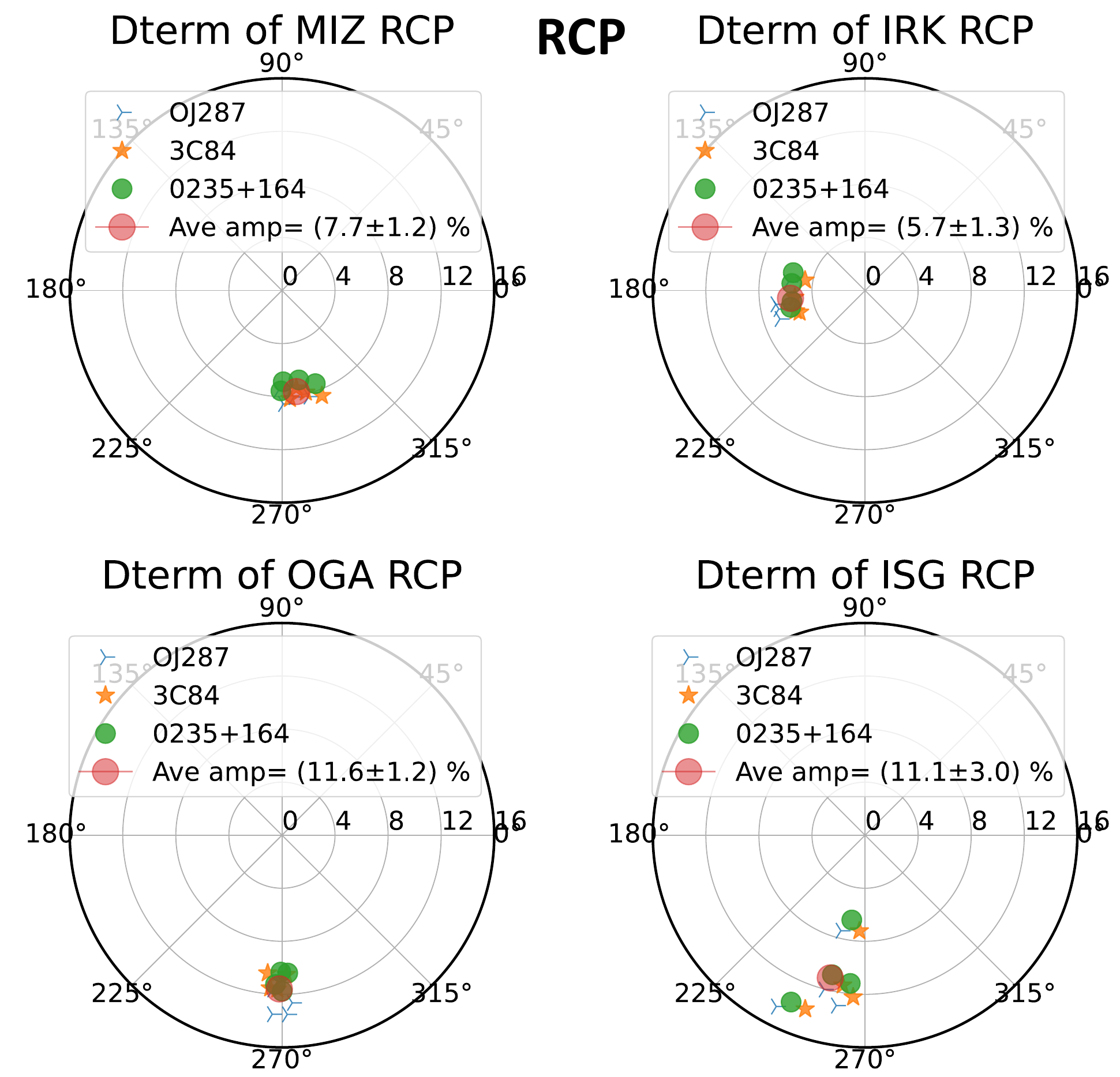}
%\end{adjustwidth}
\caption{{D-terms}  %MDPI: Please check the overlapped content
 obtained at 43 GHz. Description of what is contained in the Figure~\ref{fig5} caption. \label{fig6}}
\end{figure}

One of the primary purposes of our experiments was to estimate D-terms for each VERA telescope at 22 GHz and 43 GHz. In Figure~\ref{fig5}, we show the D-term solutions at the 22 GHz band obtained for each telescope and polarization. In each of the eight sub-panels, we plot the solutions for all the four IFs, obtained from three different calibrators OJ287, 3C84, and 0235+164, as well as a solution averaged over all the calibrators and four IF sub-bands. Figure~\ref{fig6} displays the corresponding results for the 43 GHz band. The values of D-terms for each VERA telescope are summarized in Table~\ref{tab3}.

%Figure 6

At the 22 GHz band, the D-term solutions derived from different calibrators were in very good agreement with each other at all four telescopes. The D-terms amplitudes were confirmed to be 4--7\% across the four telescopes, with a standard deviation of $\sim$1.5\%. We note that the D-term solutions seem to show some slight difference between the IF sub-bands. This could be caused by the limited SNR of our D-term derivation process, or we could be seeing a possible frequency dependence of the D-terms due to the wide  bandwidth ($\sim$2\,GHz) of our observations, which should be examined in more detail in future observations. 

At the 43 GHz band, the D-term solutions derived from different calibrators were also in good agreement with each other at all the telescopes. The D-term amplitudes at 43 GHz were somewhat larger than those at 22 GHz, but still largely constrained to be $\lesssim$~10\% (except for LCP at the Ishigaki-Jima telescope). It is uncertain whether or not this relatively larger D-term for Ishigaki was temporal. %MDPI: Is the bold necessary? YH: Removed.
 Compared with the D-terms ($<$$\sim$5\%) obtained for Very-Long-Baseline Array (VLBA) telescopes at 22 and 43 GHz \cite{gom02}, the D-terms of VERA are larger. It is not clear whether this is due to better calibration accuracy as a result of the larger number of VLBA baselines, or due to differences in the antenna feeds. Note that the D-terms of the KVN are $\sim$3.5--11\% at 22 GHz and $\sim$4.5--6.5\% at 43 GHz \cite{par18}, closer to those of VERA.  Multiepoch measurements would be necessary to further test the stability of the D-terms over time. 

Nevertheless, the D-terms results obtained from our first VERA wideband polarimetric experiments are overall very encouraging without any serious issues. This indeed allowed us to produce reliable linear polarization images with VERA, as described in the following subsection.

\vspace{-3pt}
\vspace{-3pt}  

\begin{table}[H] 
\caption{D-terms \textsuperscript{1} of the 20-m VERA telescopes (expressed in percent) \label{tab3}}
\newcolumntype{C}{>{\centering\arraybackslash}X}
\begin{tabularx}{\textwidth}{lCCCC}
\toprule 
\textbf{Telescope}	& \textbf{22 GHz, RCP}	& \textbf{22 GHz, LCP} & \textbf{43 GHz, RCP} & \textbf{43 GHz,  LCP}\\
\midrule
Mizusawa (MIZ)		& 6.3 $\pm$ 2.0			& 6.2 $\pm$ 1.7    & 7.7 $\pm$ 1.2 & 9.0 $\pm$ 2.0 \\
Iriki (IRK)		& 7.1 $\pm$ 1.3			& 7.0 $\pm$ 1.3 &         5.7 $\pm$ 1.3    & 5.3 $\pm$ 1.3 \\
Ogasawara (OGA) &  5.7 $\pm$	1.0		&5.5 $\pm$ 1.3   &     11.6 $\pm$ 1.2   & 10.9 $\pm$ 1.2     \\
Ishigak-Jima (ISG)	& 4.0 $\pm$ 1.9		& 4.6 $\pm$ 1.9  &  11.1 $\pm$ 3.0 &   14.3 $\pm$ 2.2 \\
\bottomrule
\end{tabularx}
\noindent{\footnotesize{\textsuperscript{1} Averaged over the four IF channels.}}
\end{table}
\subsection{Linear Polarization Images}

\begin{figure}[H]
  
\begin{adjustwidth}{-\extralength}{0cm}
\centering

 \begin{tabular}{cc}
\includegraphics[width=0.56\columnwidth]{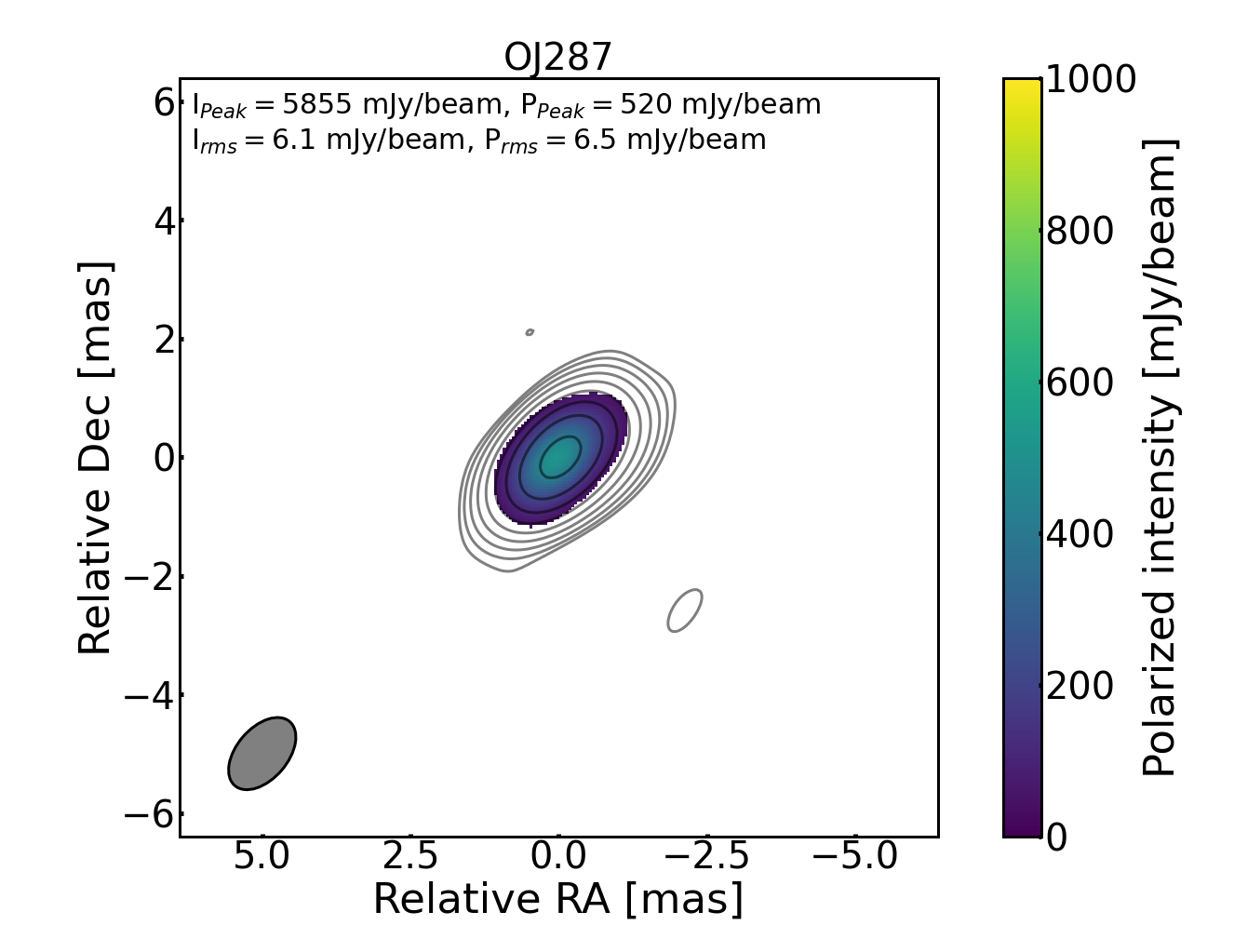}
&\includegraphics[width=0.56\columnwidth]{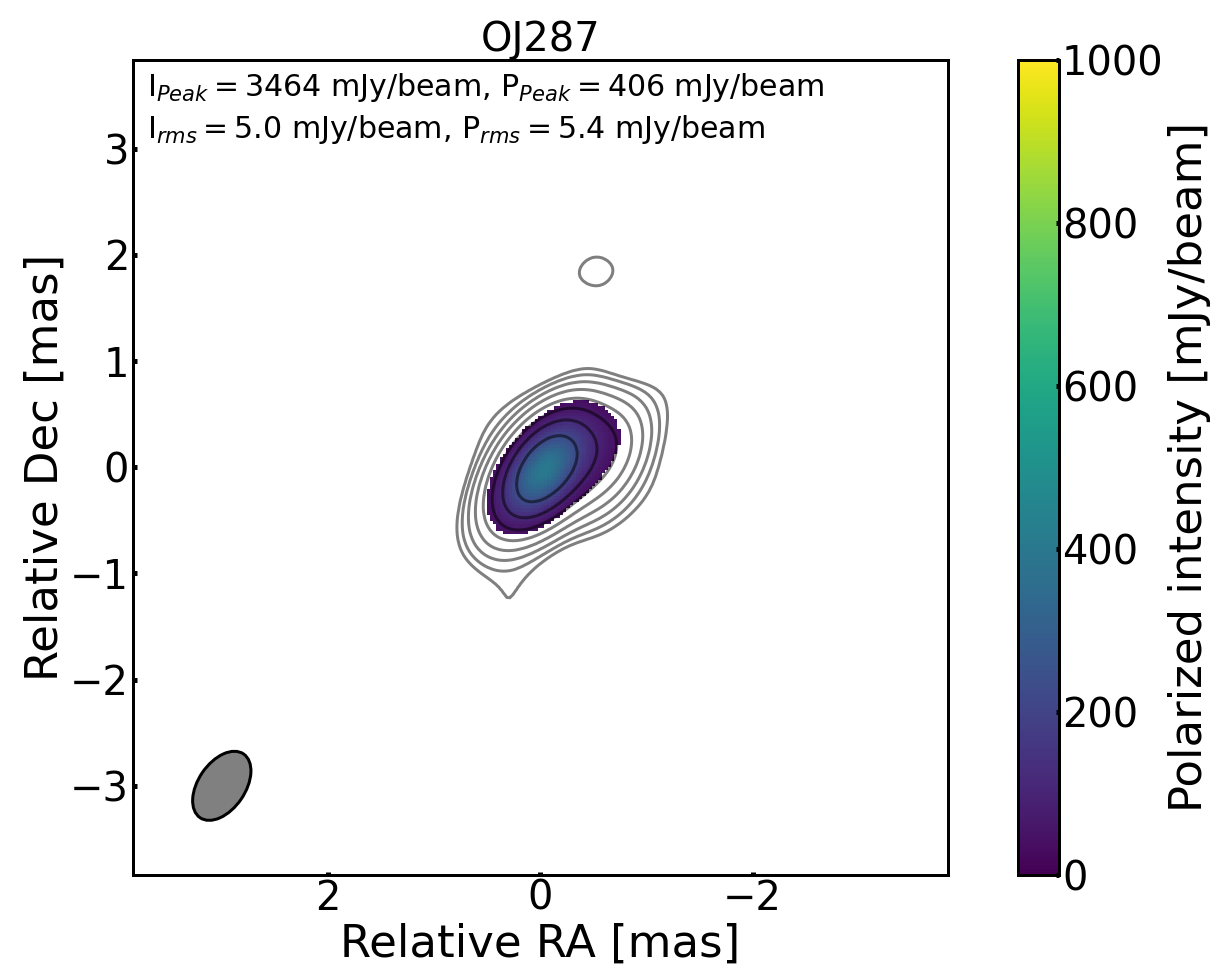}\\
\includegraphics[width=0.56\columnwidth]{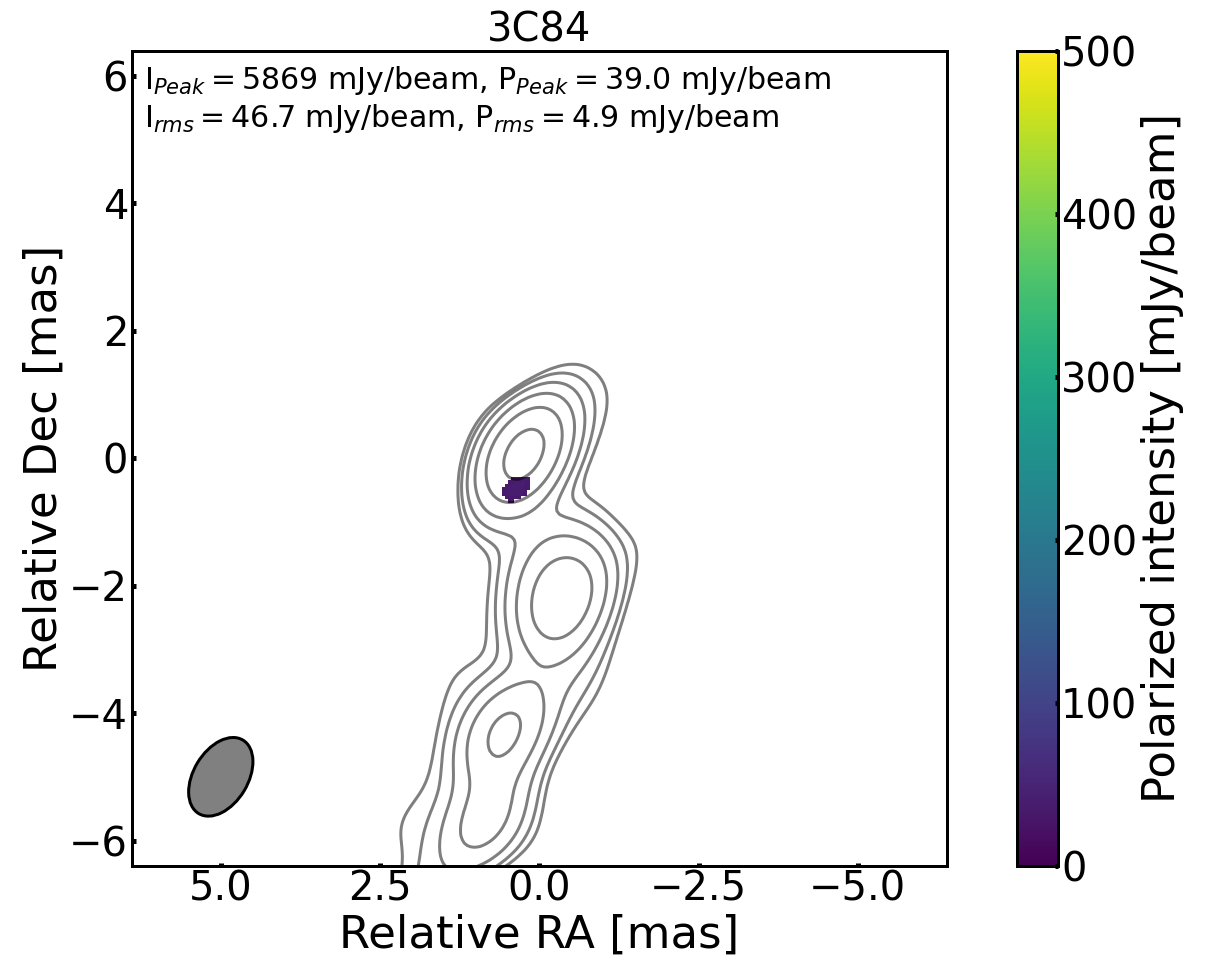}
& \includegraphics[width=0.56\columnwidth]{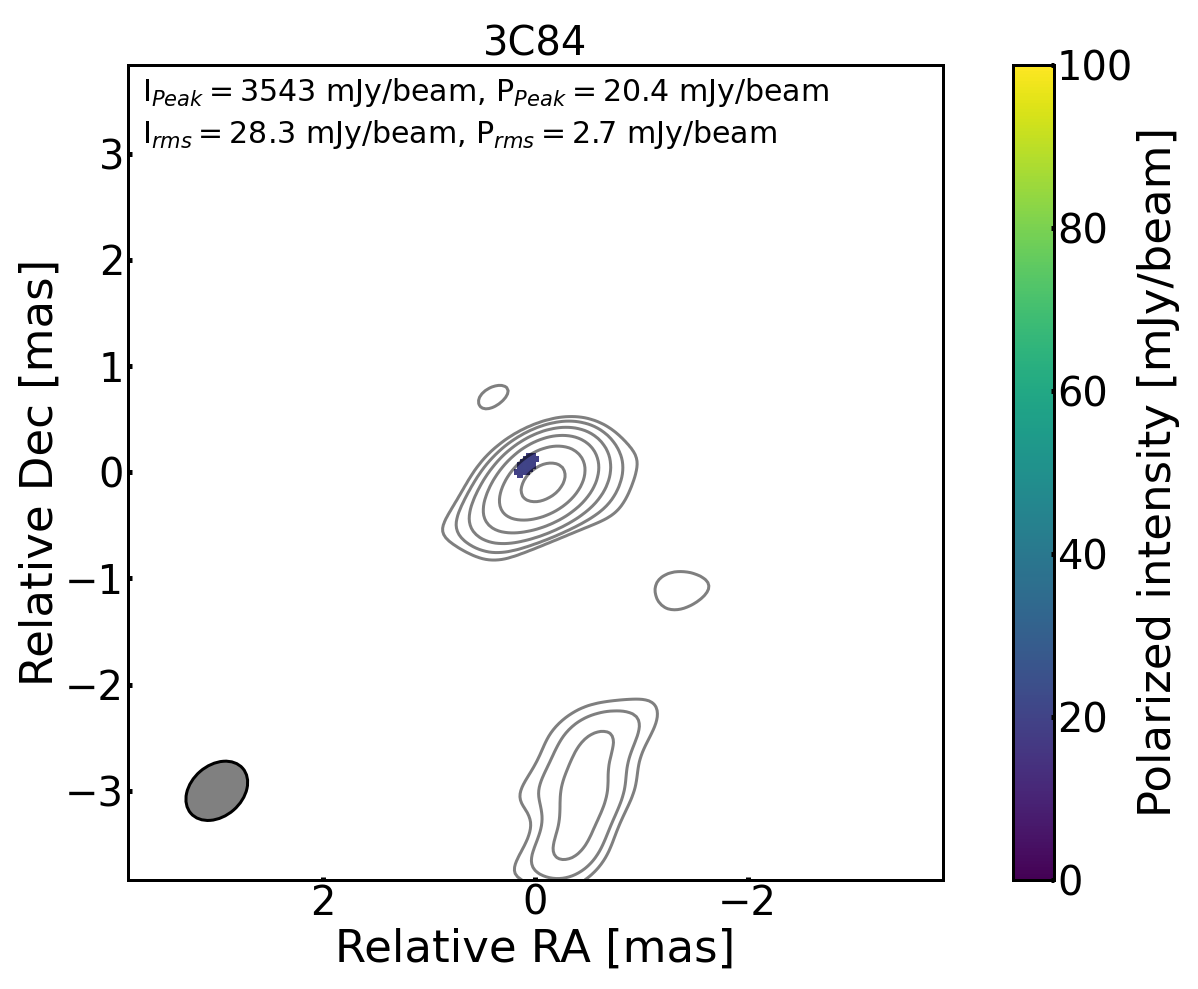}\\
  \includegraphics[width=0.56\columnwidth]{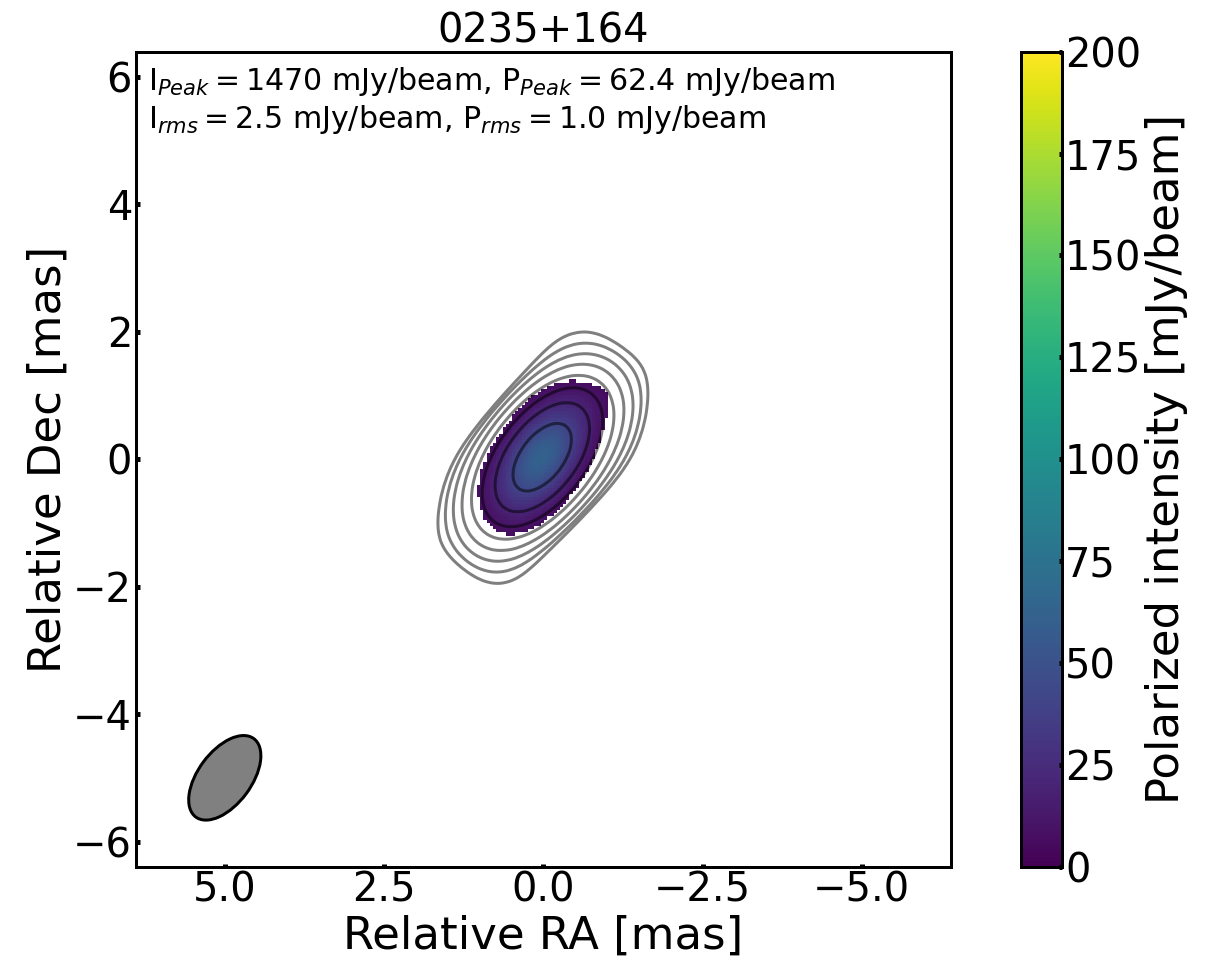}& 
        \includegraphics[width=0.56\columnwidth]{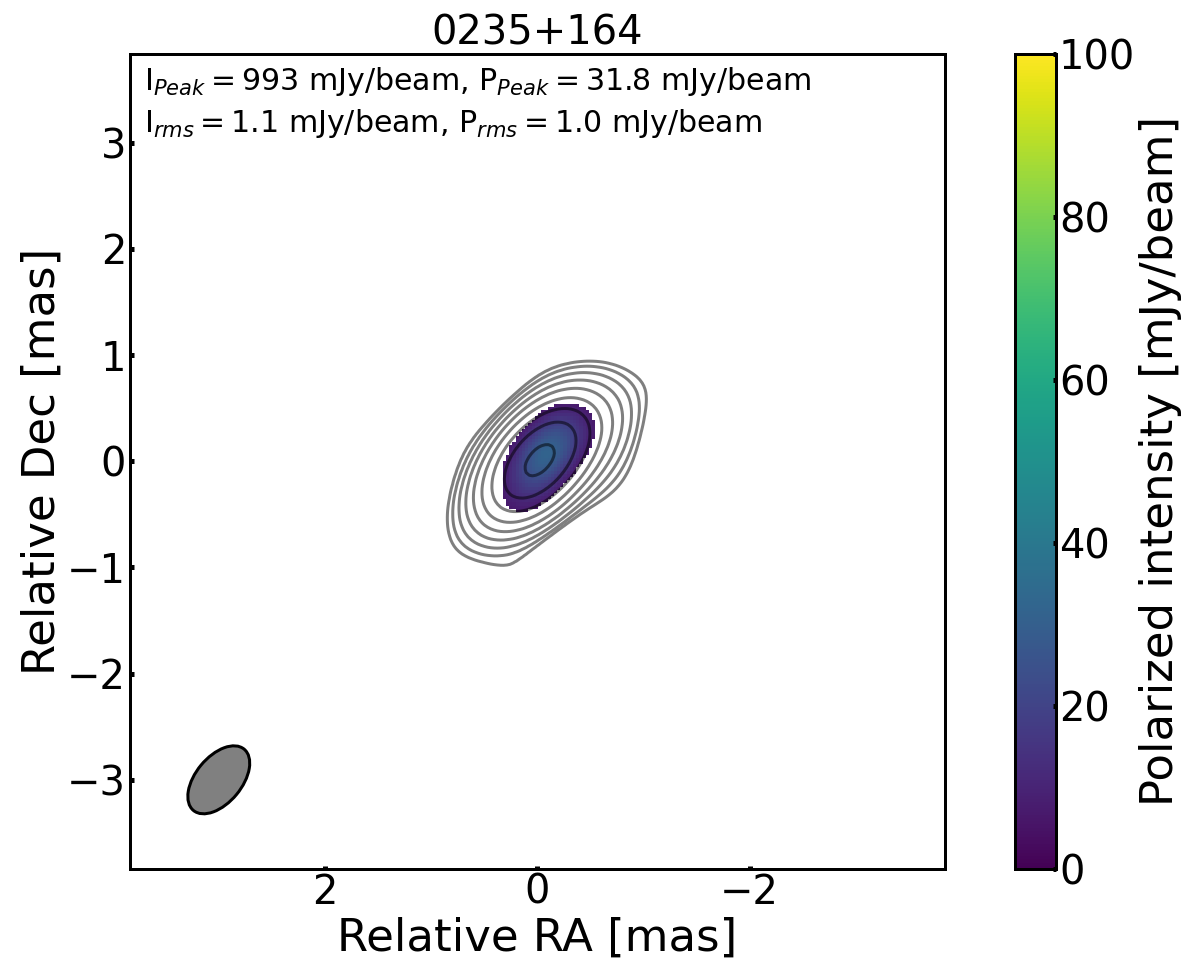}\\
\end{tabular}

%        \includegraphics[keepasectratio, width=\columnwidth]{Polflux.map.vera.OJ287.k.png}

%        \includegraphics[keepasectratio, width=\columnwidth]{Polflux.map.vera.OJ287.q.png}

%    \caption{22 GHz (\textbf{upper}) and 43 GHz  (\textbf{lower}) VERA total intensity (contour) and linearly polarized intensity (color) maps of OJ 287. In the both maps, the polarized intensity is denoted with a vertical color scale in mJy beam$^{-1}$. The synthesized beam is plotted at bottom left in each panel. \label{fig5}}
    
%        \includegraphics[keepasectratio, width=\columnwidth]{Polflux.vera.3C84.k.5sigma_202208.png}
 
%includegraphics[keepasectratio, width=\columnwidth]{Polflux.vera.3C84.q.5sigma.png}

%    \caption{22 GHz (\textbf{upper}) and 43 GHz  (\textbf{lower}) VERA total intensity (contour) and linearly polarized intensity (color) maps of 3C84. In the both maps, the polarized intensity is denoted with a vertical color scale in mJy beam$^{-1}$. The synthesized beam is plotted at bottom left in each panel. \label{fig5}}
%    \label{fig:my_label}
%\end{figure}
%\begin{figure}[htbp]

%        \includegraphics[keepasectratio, width=\columnwidth]{Polflux.vera.0235.k.20220916.png}
      
 %       \includegraphics[keepasectratio, width=\columnwidth]{Polflux.vera.0235.q.3sigma.png}
%

\end{adjustwidth}
    \caption{The 22 GHz (\textbf{left}) and 43 GHz (\textbf{right}) VERA total intensity (contour) and linearly polarized intensity (color) maps of OJ287(\textbf{top}), 3C84(\textbf{middle}), and 0235+164(\textbf{bottom}). The 1 $\sigma$ noise and peak  values of the total and polarized intensity are shown at the top of each panel. The contours are plotted with increasing powers of two, starting from the 3 $\sigma$ level. 
%%% RMC  next sentence says "both maps", but there are 6...  "both maps" -> "each map"(?) [if so, "bars"->"bar" later in the sentence]  applies only to one source's maps(?)
In both the 22 and 43 GHz maps, the polarized intensity is scaled up to 7 $\sigma$, with the color scale in mJy beam$^{-1}$ given in the vertical bars to the right. The synthesized beam is plotted at the bottom left in each panel. \label{fig7}}
%    \label{fig7}
\end{figure}

In Figure~\ref{fig7}, we present linear polarization ($P = \sqrt{Q^2 + U^2}$) images obtained for OJ287, 3C84, and 0235 + 164, which are overlaid on total intensity (Stokes I) contour maps. Images at both the 22 and 43 GHz bands are shown, and polarization emission detected at SNR $>$ 5 is shown for all images. Note that here we do not present EVPA distributions, since the correction of absolute EVPA requires external information from other facilities, which were not performed in our analysis. 
%%% RMC the flux-density numbers in this sentence don't match table.4, but the fractional degree of polarization numbers do match
For blazars OJ287 and 0235 + 164, we detected significant polarization flux densities from the core regions both at 22 and 43 GHz bands. The integrated polarized flux densities over a one mas square of the core region of OJ287 were 463 mJy and 442 mJy at 22 and 43 GHz bands, which correspond to fractional polarization degrees of 8.5\% and 10.4\%, respectively. For 0235+164, the integrated polarized flux densities over a one mas square of the core region were 54.6 mJy and 31.1 mJy at the 22 and 43 GHz bands, which correspond to fractional polarization degrees of 4.0\% and 2.9\% at 22 and 43 GHz, respectively. %MDPI: Is the bold necessary? YH: Removed.
%For blazars OJ287 and 0235+164, we detected significant polarization flux densities from the core regions both at 22 and 43 GHz bands. The recovered polarized flux densities of OJ287 were 176 mJy and 148 mJy at 22 and 43 GHz bands (at SNR 20 and 19), which correspond to a fractional polarization degree of 8.5\% and 11.2\%, respectively. For 0235+164, we recovered polarized flux densities of 30 mJy and 12 mJy at 22 and 43 GHz bands (at SNR 21 and 9), which correspond to a fractional polarization degree of 4.2\% and 3.1\%, respectively. 
%%% RMC  not sure what the very next sentence is comparing:  which are "these characteristics"?  The above sentences talk about the results of the VERA observations, so it wouldn't be making any sense to have those be "these characteristics" -- but nothing else seems to have been discussed previously...  Is it covered by the sentence "Bearing in mind..." that occurs in sect.5.3, para.1?  If so, can be omitted here.
%The polarimetric images obtained by VERA confirmed these characteristics. 
Although the nucleus of radio galaxy 3C84 is well known as a very weakly polarized source at these frequencies \cite{tay06}, we detected the polarized emission.
%we marginally detected polarized emission from it at 43 GHz despite its sufficient brightness in Stokes I (at 22 GHz, only an upper limit resulted). 
 The errors of linear polarization flux density, P$_{\rm rms}$, are estimated as the following: 
\begin{equation} 
    P_{rms} = \frac{\sigma_{Q}+\sigma_{U}}{2} ,
\end{equation} 
where $\sigma_{Q}$ and $\sigma_{U}$ are rms noises in the Stokes Q and U images, respectively.

Fractional polarization values are obtained by simply dividing the integrated polarized flux density $P_{tot}$ by the integrated total flux density $I_{tot}$. Fractional polarization errors ($\sigma_{m}$) are calculated by dividing P$_{\rm rms}$ by the peak intensity (Stokes I), as the following; see, e.g.,\citep{hov12}:
%{\bf Fractional polarization values are obtained by simply dividing the average polarized flux density of a region by the average the total intensity flux density. Fractional polarization errors ($\sigma_{m}$) are calculated by dividing P$_{\rm rms}$ by the total intensity (Stokes I) as the following, see, e.g.,\citep{hov12}:}
\begin{equation} 
    \sigma_{m} = \frac{P_{rms}}{I_{Peak}} 
\end{equation} 
%\begin{equation} 
%    \sigma_{m} = \frac{P_{rms}}{I} 
%\end{equation} 
We list the parameters of total flux densities, polarized flux densities, fractional linear polarization values (average values of the fractional polarization of an area indicated by color in each panel), and rms values for each map in Table~\ref{tab4}.

\vspace{-5pt} 
\begin{table}[H] 
\caption{Summary of 22 and 43 GHz Image Parameters \label{tab4}}

\begin{adjustwidth}{-\extralength}{0cm}
\centering 
\setlength{\cellWidtha}{\fulllength/8-2\tabcolsep-0.2in}
\setlength{\cellWidthb}{\fulllength/8-2\tabcolsep-0.1in}
\setlength{\cellWidthc}{\fulllength/8-2\tabcolsep-0.1in}
\setlength{\cellWidthd}{\fulllength/8-2\tabcolsep-0.0in}
\setlength{\cellWidthe}{\fulllength/8-2\tabcolsep+0.1in}
\setlength{\cellWidthf}{\fulllength/8-2\tabcolsep+0.1in}
\setlength{\cellWidthg}{\fulllength/8-2\tabcolsep+0.1in}
\setlength{\cellWidthh}{\fulllength/8-2\tabcolsep+0.1in}
\scalebox{1}[1]{\begin{tabularx}{\fulllength}{
>{\PreserveBackslash\centering}m{\cellWidtha}
>{\PreserveBackslash\centering}m{\cellWidthb}
>{\PreserveBackslash\centering}m{\cellWidthc}
>{\PreserveBackslash\centering}m{\cellWidthd}
>{\PreserveBackslash\centering}m{\cellWidthe}
>{\PreserveBackslash\centering}m{\cellWidthf}
>{\PreserveBackslash\centering}m{\cellWidthg}
>{\PreserveBackslash\centering}m{\cellWidthh}}
\toprule 
\textbf{Object}	&\textbf{Band} &\boldmath{$I_{tot}$}\textbf{ \textsuperscript{1}} & \boldmath{$P_{tot}$}\textbf{\textsuperscript{2}} &\boldmath{$I_{peak}$}\textbf{\textsuperscript{3}}& \boldmath{$I_{rms}$}\textbf{\textsuperscript{4}}&\boldmath{$P_{rms}$}\textbf{\textsuperscript{5}} 	& \textbf{R}\boldmath{$_{pol}$}\textbf{\textsuperscript{6}}  \\
	&\textbf{(GHz)}&\textbf{(Jy)}&\textbf{(mJy)}&\textbf{(Jy beam}\boldmath{$^{-1}$}\textbf{)}& \textbf{(mJy beam}\boldmath{$^{-1}$}\textbf{)} &\textbf{(mJy beam}\boldmath{$^{-1}$}\textbf{)}&\textbf{ (\%)}\\
\midrule
%OJ287 &22&5.4&463.3&5855&6.1&6.5  &8.5$\pm$0.1 \\
OJ287 &22&5.4&463.3&5.9&6.1&6.5  &8.5 $\pm$ 0.1 \\
%OJ287 &43&4.2&441.7&3464&5.0&5.4   & 10.4$\pm$0.2 \\
OJ287 &43&4.2&441.7&3.5&5.0&5.4   & 10.4 $\pm$ 0.2 \\
%OJ287 &22&5.6& 478&6.1&6.5  &8.5 $\pm$0.29 \\
%OJ287 &43&3.5&394&5.0&5.4   & 11.2 $\pm$ 0.41 \\
%
%3C 84 &22&5.5&15.4 &5869&46.7 &4.9   & 0.3$\pm$0.1\\
%3C 84 &43&4.3&16.0&3543&28.2  &2.7   & 0.4$\pm$0.1\\
3C 84 &22&5.5&15.4 &5.9&46.7 &4.9   & 0.3 $\pm$ 0.1\\
3C 84 &43&4.3&16.0&3.5&28.2  &2.7   & 0.4 $\pm$ 0.1\\
%
%0235+164 &22&1.4&54.6 &1470& 2.5	&1.0   & 4.0$\pm$ 0.1\\
%0235+164 &43&  1.1& 31.1&993&   1.1&1.0   & 2.9$\pm$0.1 \\
0235+164 &22&1.4&54.6 &1.5& 2.5	&1.0   & 4.0 $ \pm $ 0.1\\
0235+164 &43&  1.1& 31.1&1.0&   1.1&1.0   & 2.9 $\pm$ 0.1 \\
%0235+164 &22&1.35&55 & 2.5	&1   & 4.2 $\pm$ 0.17\\
%0235+164 &43&  0.7& 22.4&   1.1&1   & 3.1 $\pm$0.2 \\
%
%3C 84 &22&0.34&3.7&46.7  &4.9   & 1.1 $\pm$0.15\\
%3C 84 &43&0.7&4.4&28.2  &2.7   & 0.6$\pm$ 0.14\\
\bottomrule
\end{tabularx}}
\end{adjustwidth}
%\noindent{\footnotesize{\textsuperscript{4} Linear polarization rate error }}
\noindent{\footnotesize{\textsuperscript{1} Total flux (Stokes I) density }}
\noindent{\footnotesize{\textsuperscript{2} Total linear polarization flux density }}
\noindent{\footnotesize{\textsuperscript{3} Peak flux density }}
\noindent{\footnotesize{\textsuperscript{4} Stokes I image rms }}
\noindent{\footnotesize{\mbox{\textsuperscript{5} Polarized image rms} }}
\noindent{\footnotesize{\textsuperscript{6} Fractional linear polarization}.}

\end{table}

\subsection{Verification}
Given the relatively small D-term values, we conclude that we can conduct polarimetric VLBI observations at both 22 and 43 GHz using VERA.
All of the three sources discussed in the previous subsection are regularly monitored with VLBA at 15\,GHz through the MOJAVE program~\cite{lister2018} and at 43\,GHz through the Boston University Blazar program~\citep{jor17}. Bearing in mind the different times of observations, angular resolutions, and observing frequencies, the polarization characteristics of these sources obtained with VERA are consistent overall with those typically measured from the VLBA observations, confirming the capability of VERA for polarimetric study.
A rotation measure of the narrow-line Seyfert 1 galaxy was tentatively obtained from our data (Takamura et al., in preparation), which demonstrates that the detection of large rotation measure can be made by the combination of the new dual-polarization and ultrawideband ($\sim$2GHz) capabilities of VERA. 

\section{Future Prospects}
First of all, we aim to present scientific results obtained from the test observations of Faraday rotation measurement towards narrow-line Seyfert 1 galaxies in March 2022, and commissioning that will be conducted in the near future (Takamura et al., in preparation). After commissioning is completed, for example, an extensive mapping observation campaign of narrow-line Seyfert 1 galaxies will begin utilizing the ultrawideband polarimetric VLBI capability. 

Additional scientific capabilities will be implemented to both the front- and back-end in the future. For example, simultaneous observations at 22 and 43 GHz with dual polarizations at 16 \gbps will be enabled, which will significantly improve the sensitivity of measuring spectral indices of active galactic nuclei. Moreover, "super" ultrawideband observation at 32 \gbps will be possible in the future.  % Please check intended meaning is retained.

We started VLBI polarimetric test observations with the three 21 m telescopes of the Korean VLBI Network (KVN).%MDPI: Is the bold necessary? => YH: No, removed again.
%, providing better u-v coverage and sensitivity of polarimetric mapping observations at 22 and 43 GHz.
{With the KVN, both single-dish and VLBI monitoring observations of the polarization of Blazars at 22 and 43 GHz are being conducted.  This will facilitate EVPA calibration for polarimetric observations using VERA, or with a combined array observation with the KVN and VERA (KaVA) in the future; see, e.g., \cite{kan15}.  Such KaVA capabilities would provide better UV coverage and sensitivity for polarimetric mapping observations at 22 and 43 GHz.} % Please check intended meaning is retained.
%
%%%%%%%%%%%%%%%%%%%%%%%%%%%%%%%%%%%%%%%%%%
\section{Summary}
The new dual-circular polarization receiving system was designed, and the installation of the new RCP receivers at all four telescopes of VERA was completed in August 2019.
With the new receivers, the new VERA front-end enables VLBI polarimetry, and we aim to combine the VERA with KVN in the near future for more sensitive polarimetric imaging.
The new back-end can process and record data at 16 \gbps,  
 providing up to factor of four improvement in sensitivity compared with the earlier capability.
 
In March 2022, the first VERA polarimetric observations at 22 and 43 GHz were tested using the ultrawideband recording system at 16 \gbps (8 \gbps per polarization.%MDPI: Is the bold necessary?  YH: No, removed.
 With the current front-end configuration, the recording rate can be extended up to 32 \gbps. 
%The ultra-wide band observation needs the recording rate expansion of other VLBI stations or arrays, such as KVN.}
Exploitation of the ultrawideband mode in more extensive VLBI arrays, such as KaVA, might need a comparable recording rate enhancement at other participating telescopes or arrays.

From these observations, we obtained for the first time 22 and 43 GHz total intensity and linear polarization maps of extragalactic sources using the VERA.
Thus, the new ultrawideband recording has enabled sensitive polarization observations by VERA. We have no doubt that these developments are now, and will be in the future, of great importance for VLBI collaborations, primarily in the East Asian region.
%%%%%%%%%%%%%%%%%%%%%%%%%%%%%%%%%%%%%%%%%%
%\section{Patents}

%This section is not mandatory, but may be added if there are patents resulting from the work reported in this manuscript.

%%%%%%%%%%%%%%%%%%%%%%%%%%%%%%%%%%%%%%%%%%
\vspace{6pt} 

%%%%%%%%%%%%%%%%%%%%%%%%%%%%%%%%%%%%%%%%%%
%% optional
%\supplementary{The following supporting information can be downloaded at:  \linksupplementary{s1}, Figure S1: title; Table S1: title; Video S1: title.}


%%%%%%%%%%%%%%%%%,%%%%%%%%%%%%%%%%%%%%%%%%
\authorcontributions{Project management, Y.H. and K.H.; observations, K.H. and M.T.; instrumentation, S.S., T.O., Y.H., and A.Y.; software, T.O.; data analysis, M.T. and K.H.; writing and editing, Y.H.; funding acquisition, Y.H. and K.H. All authors have read and agreed to the published version of \mbox{the manuscript.}}

%The following statements should be used ``Conceptualization, X.X. and Y.Y.; methodology, X.X.; software, X.X.; validation, X.X., Y.Y. and Z.Z.; formal analysis, X.X.; investigation, X.X.; resources, X.X.; data curation, X.X.; writing---original draft preparation, X.X.; writing---review and editing, X.X.; visualization, X.X.; supervision, X.X.; project administration, X.X.; funding acquisition, Y.Y.
% \href{http://img.mdpi.org/data/contributor-role-instruction.pdf}{CRediT taxonomy} for the term explanation. Authorship must be limited to those who have contributed substantially to the work~reported.}

\funding{This research was funded by JSPS KAKENHI grant number JP15H03644 (PI:Y.Hagiwara) and JP19H01943 (PI:K.Hada). }
% \url{https://search.crossref.org/funding}, any errors may affect your future funding.}

\institutionalreview{{Not applicable}} %MDPI:In this section, you should add the Institutional Review Board Statement and approval number, if relevant to your study. You might choose to exclude this statement if the study did not require ethical approval. Please note that the Editorial Office might ask you for further information. Please add “The study was conducted in accordance with the Declaration of Helsinki, and approved by the Institutional Review Board (or Ethics Committee) of NAME OF INSTITUTE (protocol code XXX and date of approval).” for studies involving humans. OR “The animal study protocol was approved by the Institutional Review Board (or Ethics Committee) of NAME OF INSTITUTE (protocol code XXX and date of approval).” for studies involving animals. OR “Ethical review and approval were waived for this study due to REASON (please provide a detailed justification).” OR “Not applicable” for studies not involving humans or animals.}

\informedconsent{Not applicable} %MDPI:Any research article describing a study involving humans should contain this statement. Please add ``Informed consent was obtained from all subjects involved in the study.'' OR ``Patient consent was waived due to REASON (please provide a detailed justification).'' OR ``Not applicable'' for studies not involving humans. You might also choose to exclude this statement if the study did not involve humans.

\dataavailability{Not applicable} %MDPI:In this section, please provide details regarding where data supporting reported results can be found, including links to publicly archived datasets analyzed or generated during the study. Please refer to suggested Data Availability Statements in section ``MDPI Research Data Policies'' at \url{https://www.mdpi.com/ethics}. If the study did not report any data, you might add ``Not applicable'' here.

%\informedconsent{Any research article describing a study involving humans should contain this statement. Please add ``Informed consent was obtained from all subjects involved in the study.'' OR ``Patient consent was waived due to REASON (please provide a detailed justification).'' OR ``Not applicable'' for studies not involving humans. You might also choose to exclude this statement if the study did not involve humans.

%Written informed consent for publication must be obtained from participating patients who can be identified (including by the patients themselves). Please state ``Written informed consent has been obtained from the patient(s) to publish this paper'' if applicable.}
%\dataavailability{Not applicable}
%\dataavailability{In this section, please provide details regarding where data supporting reported results can be found, including links to publicly archived datasets analyzed or generated during the study. Please refer to suggested Data Availability Statements in section ``MDPI Research Data Policies'' at \url{https://www.mdpi.com/ethics}. If the study did not report any data, you might add ``Not applicable'' here.} 

\acknowledgments{We gratefully acknowledge the support of the Mizusawa VLBI Observatory staff for their assistance with the observing. YH thanks Noriyuki Kawaguchi for his useful advice on the development of the VERA front-end system and also thanks Takeshi Ohno for his efforts during the receiver installation process. The authors thank Bob Campbell for comments and suggestions that significantly improved the manuscript. VERA is operated by National Astronomical Observatory \mbox{of Japan.} Part of the research was supported by the Inoue Enryo Memorial Grant, TOYO University in Japan.}

%\conflictsofinterest{The authors declare no conflict of interest.}

%Authors must identify and declare any personal circumstances or interest that may be perceived as inappropriately influencing the representation or interpretation of reported research results. Any role of the funders in the design of the study; in the collection, analyses or interpretation of data; in the writing of the manuscript, or in the decision to publish the results must be declared in this section. If there is no role, please state ``The funders had no role in the design of the study; in the collection, analyses, or interpretation of data; in the writing of the manuscript, or in the decision to publish the~results''.} 

\conflictsofinterest{The authors declare no conflict of interest.} %MDPIDeclare conflicts of interest or state ``The authors declare no conflict of interest.'' Authors must identify and declare any personal circumstances or interest that may be perceived as inappropriately influencing the representation or interpretation of reported research results. Any role of the funders in the design of the study; in the collection, analyses or interpretation of data; in the writing of the manuscript; or in the decision to publish the results must be declared in this section. If there is no role, please state ``The funders had no role in the design of the study; in the collection, analyses, or interpretation of data; in the writing of the manuscript; or in the decision to publish the~results''.

%% Optional
%\sampleavailability{Samples of the compounds ... are available from the authors.}

%%%%%%%%%%%%%%%%%%%%%%%%%%%%%%%%%%%%%%%%%%
%% Only for journal Encyclopedia
%\entrylink{The Link to this entry published on the encyclopedia platform.}

%%%%%%%%%%%%%%%%%%%%%%%%%%%%%%%%%%%%%%%%%%
%% Optional
\abbreviations{Abbreviations}{
The following abbreviations are used in this manuscript:\\

%19H01943
\noindent 
\begin{tabular}{@{}ll}
AGN & Active Galactic Nuclei \\
EVPA & Electric Vector Position Angle \\
KaVA & KVN and VERA \\
KVN & Korean VLBI Network \\
LCP/RCP & Light-/Right-Handed Circular Polarization\\
LNA & Low-noise amplifiers \\
MOJAVE &Monitoring Of Jets in Active galactic nuclei with VLBA Experiments\\
OCTAD&Optically Connected Array for VLBI Exploration Analog-to-Digital Converter\\
VERA & VLBI Exploration of Radio Astrometry\\
VLBI & Very-Long-Baseline Interferometry
%MDPI & Multidisciplinary Digital Publishing Institute\\
%LD & Linear dichroism
\end{tabular}}

%%%%%%%%%%%%%%%%%%%%%%%%%%%%%%%%%%%%%%%%%%
%% Optional
%\appendixtitles{no} % Leave argument "no" if all appendix headings stay EMPTY (then no dot is printed after "Appendix A"). If the appendix sections contain a heading then change the argument to "yes".
%\appendixstart
%\appendix
%\section[\appendixname~\thesection]{}
%\subsection[\appendixname~\thesubsection]{}
%The appendix is an optional section that can contain details and data supplemental to the main text---for example, explanations of experimental details that would disrupt the flow of the main text but nonetheless remain crucial to understanding and reproducing the research shown; figures of replicates for experiments of which representative data are shown in the main text can be added here if brief, or as Supplementary Data. Mathematical proofs of results not central to the paper can be added as an appendix.

%\section[\appendixname~\thesection]{}
%All appendix sections must be cited in the main text. In the appendices, Figures, Tables, etc. should be labeled, starting with ``A''---e.g., Figure A1, %Figure A2, etc.
%%%%%%%%%%%%%%%%%%%%%%%%%%%%%%%%%%%%%%%%%%
\begin{adjustwidth}{-\extralength}{0cm}

\reftitle{References}

\end{adjustwidth}
\end{document}